\documentclass[aps,prx,twocolumn,floatfix,amssymb]{revtex4-2}

\usepackage{graphicx,dcolumn,bm,xstring,braket,amsmath,comment,placeins}
\usepackage{color}
\usepackage[colorlinks,allcolors=blue,bookmarks=true]{hyperref}

\graphicspath{{./Figs/}} 

\newcommand{\trc}{\mbox{trace}}

\newcommand{\amatrix}[1]{\begin{matrix} #1 \end{matrix}}

\renewcommand{\ket}[1]{\left| #1 \right\rangle}
\renewcommand{\braket}[1]{\left\langle #1 \right\rangle }

\renewcommand{\Braket}[2]{\left\langle #1 \middle| #2 \right\rangle}
\newcommand{\BraKet}[3]{\left\langle #1 \middle| #2 \middle| #3 \right\rangle}

\newcommand{\beq}{\begin{eqnarray}}
\newcommand{\eeq}{\end{eqnarray}} 

\newcommand{\hide}[1]{}  

\newcommand{\TODO}[1]{} 
 
\newcommand{\Eq}[1]{{\textcolor{blue}{Eq.}}~\!\!(\ref{#1})} 
\newcommand{\Sec}[1]{{\textcolor{blue}{Sec.}}~(\ref{#1})} 
\newcommand{\App}[1]{{\textcolor{blue}{Appendix}}~\ref{#1}} 
\newcommand{\Fig}[1] {{\textcolor{blue}{Fig.}}~\!\!\ref{#1}}
\newcommand{\sect}[1]{{\bf #1.-- }}

\newcommand{\hrefl}[2]{\href{#2}{(#1)}}

\begin{document}
\title{Quantum tomography of the superfluid-insulator transition \\ for a mesoscopic atomtronic ring} 

\author{Yehoshua Winsten, Doron Cohen} 

\affiliation{
\mbox{Department of Physics, Ben-Gurion University of the Negev, Beer-Sheva 84105, Israel} 
}

\begin{abstract}
We provide a phase-space perspective for the analysis of the superfluid-insulator transition for finite-size Bose-Hubbard circuits. We explore how the eigenstates parametrically evolve as the inter-particle interaction is varied, paying attention to the fingerprints of chaos at the quantum phase-transition. Consequently, we demonstrate that the tomographic spectrum reflects the existence of mixed-regions of chaos and quasi-regular motion in phase-space. This tomographic semiclassical approach is much more efficient and informative compared to the traditional ``level statistics" inspection. Of particular interest is the characterization of the fluctuations that are exhibited by the many-body eigenstates. In this context, we associate with each eigenstate a Higgs measure for the identification of amplitude modes of the order-parameter. Finally we focus on the formation of the lowest Goldstone and Higgs bands.  
\end{abstract}

\maketitle

\section{Introduction}
\label{s1}

The study of quantum phase transitions has long been a cornerstone of condensed matter physics. Among these transitions, the superfluid (SF) to Mott insulator (MI) transition in the Bose-Hubbard model (BHM) stands out as a paradigmatic example \cite{Leggett,BHHqpt,BHH1,BHH2}. This model provides a simplified description of interacting bosonic particles confined in a lattice potential. It captures essential themes, including super-fluidity, self-trapping, and the formation of solitons.

\sect{Quantum chaos}
If the inter-particle interaction is not too strong, the {\em ground state} of the Bose-Hubbard model exhibits superfluidity, characterized by the coherent flow of particles throughout the lattice. As the density or the strength of the interactions is increased, the system undergoes the SF-to-MI transition \cite{MISF1,MISF2,MISF3}. In fact this transition is apparent also in the parametric evolution of the {\em excited states}.  In this context, quantum chaos is a fascinating aspect that should not be overlooked \cite{dys2,kolovskiPRL,trimerKottos,BHHchains-Henn,MixedChaos,sfc,sfa,BHHtrimerChaos,BHHchainChaos1,BHHchainChaos2}. It refers to the fingerprints of chaotic dynamics that characterizes the corresponding classical model, which is the discrete nonlinear Schrodinger equation (DNLSE), aka the discrete Gross-Pitaevskii equation.

\sect{The BHM chain}
The 1D version of the BHM concerns $N$ bosons that can hope along a chain that consists of $L$ sites. For ${L>2}$ it is more illuminating to assume periodic boundary conditions (ring geometry), aka atomtronic circuit \cite{atomtronics}. The hopping frequency is~$-J$, and the on site interaction is~$U$. If one regards the Hamiltonian as a classical system that consists of $L$ coupled non-linear oscillators, the generated equation of motion is the DNLSE, which contains a single dimensionless parameter~$u$.  Upon quantization $1/N$ plays the role of Planck constant, and a second dimensionless parameter~$\gamma$ is defined.  Namely,    
\beq
u &=& \frac{NU}{J}  \ = \ \text{classical parameter} 
\\ \label{eGamma}
\gamma &=& \frac{L}{N^2}u \ = \ \text{quantum parameter}
\eeq
The classical parameter $u$ controls the appearance of self-trapped states, and the stability of superflow \cite{carr,niu,anamaria,sfa}, while the quantum parameter $\gamma$ controls the SF-MI transition. In the MI phase, it becomes important whether ${\bar{n} \equiv N/L}$ is integer or not. Thus, as $u$ is increased, the ground state changes from coherent condensation in a momentum orbital, to a fragmented site-occupation. Formally speaking, the transition is abrupt only for an infinite chain, but its mesoscopic version is clearly apparent even for a two-site model (dimer), as pointed out long ago by Leggett \cite{Leggett}, who called it a transition from Josephson-regime to Fock-regime. See also \cite{csd,DimerQPT}.
It is therefore natural to ask how the transition is reflected in a mesoscopic ring, say a trimer that has ${L=3}$ sites. The new ingredient is {\em chaos}. The question arises how the SF-MI transition is mediated or affected by chaos. 

\sect{Puzzle}
There is an extensive body of  literature about the quantum spectrum of mesoscopic Bose-Hubbard rings, and in particular about mesoscopic superfluidity. But there is a missing bridge to themes that are related to the SF-MI transition, as $u$ is increased. In particular one wonders what is the relation between the observed quantum spectrum and the prediction of the Gutzwiller Mean Field Theory (GMFT) \cite{GMFTr1,GMFTr2,GMFTr3,GMFT1,GMFT2,GMFT3}, aka slaved-bosons formalism. What is the nature of the low lying excitations \cite{Higgs,GMFTe}, and what is the border between the GMFT quasi-regular regime and the possibly ergodic quantum chaos regime.        

\sect{Strategy}
We use the term {\em quantum phase space tomography} \cite{csf} in order to emphasize that the inspection of the spectrum is not limited to ``level statistics", as e.g. in \cite{BHHchainChaos1,BHHchainChaos2}, but rather oriented to reveal the detailed relation to the underlying phase space structures, that serve as a classical skeleton for the quantum eigenstates.   
A useful strategy is to provide 3D images of the spectrum. Each point in such image represents an eigenstate, whose  vertical position is the energy. The extra dimensions (horizontal axis/axes and/or color-code) are exploited to reveal properties of the eigenstates. Such technique has been exploited e.g. to highlight Monodromy-related features in the spectrum \cite{bhm} (and further references therein). This can be complemented by inspection of representative eigenstates, using a quantum Poincare sections, as in \cite{sst}. The big picture can be summarized by a ${(u,E)}$ phase-diagram that shows the dependence of the spectrum on the the dimensionless interaction parameter~$u$.

\sect{Outline} 
In \Sec{s2} we introduce the BHM Hamiltonian, with focus on the $L{=}2$ dimer and the $L{=}3$ trimer. The main measures for characterization of eigenstates are presented in \Sec{s3}. Then we present the tomography of the spectrum for the dimer in \Sec{s4}, and proceed with the phase-diagram of the trimer in \Sec{s5}. The tomography of the spectrum and the fingerprints of chaos are discussed in \Sec{s6}. The characterization of fluctuations is worked out in \Sec{s7}, with associated results for the Higgs measure in \Sec{s9}. Finally we zoom into the lowest excitation bands in \Sec{s10}. Further discussion of open issues is presented in the summarizing \Sec{s11}. Some technical sections appear as appendices.

\section{The BHM Hamiltonian}
\label{s2}

The BHM Hamiltonian for an $L$ site system is 
\beq \label{eBHM}
&& \mathcal{H}_{\text{BHM}} \ = \ 
\frac{U}{2} \sum_{j=1}^{L} a^{\dag}_ja^{\dag}_ja_ja_j  
\nonumber \\ 
&& \ \ \ \ 
-\frac{J}{2} \sum_{j=1}^{L} 
\left( e^{i(\Phi/L)} a^{\dag}_{j{+}1}a_j + e^{-i(\Phi/L)} a^{\dag}_ja_{j{+}1}
\right)
\eeq
where $a_j$ and $a_j^{\dag}$ are annihilation and creation operators, and the summation is over the site index~$j$ modulo~$L$. 
There is a possibility to consider a ring in a rotating frame \cite{carr,sfc,sfa}. The Sagnac phase $\Phi$ is proportional to the rotation velocity, and equals zero unless we state otherwise. 
The classical version of the Hamiltonian can be written in action-angle variables. Namely, with the substitution ${a_j\mapsto \sqrt{n_j}\exp(i\varphi_j)}$ one obtains: 
\beq \label{eBHMc}
&& \mathcal{H}_{\text{BHM}}^{cl} \ = \   
\frac{U}{2}\sum_j (n_j{-}1)n_j 
\nonumber \\ 
&& \ \ \ \ 
- J \sum_{j=1}^{L} \sqrt{n_{j{+}1} n_j}
\cos((\varphi_{j{+}1}-\varphi_j) - \Phi) 
\eeq
Due to conservation of particles, ${N=\sum n_j}$ is constant of motion, and therefore we have reduction to $d=(L{-}1)$ degrees of freedom. For example, for the trimer ($L=3$) we can regard ${(n_1,n_2)}$ as ``position" space, with conjugate coordinates ${\tilde{\varphi}_1=\varphi_1-\varphi_3}$ and ${\tilde{\varphi}_2=\varphi_2-\varphi_3}$.

Optionally, one can define creation operators in momentum orbitals, namely, 
\beq
b_k^{\dagger} \ \ = \ \ \frac{1}{\sqrt{L}} \sum_j  e^{ikx_j} a_j^{\dagger}
\eeq
where ${x_j=j}$, and ${ k = (2\pi/L) \times \text{integer} }$ is defined modulo~$2\pi$. The associated occupation operators are ${\hat{n}_k=b_k^{\dagger}b_k}$.
The distinction between site-basis and momentum-basis coordinates is implied by the index ($i,j$ for sites, $k$ for momentum). 
Dropping a constant, the BHM Hamiltonian takes the following form 
\beq
\mathcal{H}_{\text{BHM}} &=& \sum_k  \left[\varepsilon_k \hat{n}_k 
-\frac{U}{2L}\hat{n}_k^2\right] 
\nonumber \\
&& + \ \text{interaction-induced-transitions}
\eeq
where the orbital energies are 
${\varepsilon_k =-J\cos(k-(\Phi/L))}$. Note that the interaction term favors condensation in momentum orbitals, which is complementary to \Eq{eBHMc} where it favors fragmented site-occupation.

The dimer Hamiltonian ($L=2$) can be written using generators of spin-rotations.  
The observable $S_z$ is defined as half the occupation difference in the site representation, namely ${S_z=\frac{1}{2}[a^{\dag}_1 a_1-a^{\dag}_2 a_2]}$. 
Then, one defines ${S_{+}=a^{\dag}_1 a_2}$ and ${S_{-} = S_{+}^{\dag}}$, 
and the associated $S_x$ and $S_y$ operators.
Accordingly, $S_x$ is identified as half the occupation difference in the momentum representation, where the momentum states are the mirror-symmetric orbitals: for the lower one is even, and the upper one is odd.  With the above notations the dimer Hamiltonian takes the following form, 
\beq  \label{eHdimer}
\mathcal{H}_{\text{dimer}} =  \left[\left(\frac{N}{2}{+}1\right)  \frac{N}{2}U\right] \ + U S_z^2 - J S_x 
\eeq
For $L>2$ ring we can generalize the spin-style notations. In particular, we can define jump operator as ${S_{i,j}=a_i^{\dagger}a_j}$, and associated $S_x$ and $S_y$ operators for each pair of sites. 

\section{Characterization of eigenstates}
\label{s3}

The eigenstates of the BHM are ordered by energy and indexed by ${\nu=0,1,2,\cdots,\mathcal{N}{-}1}$, where $\mathcal{N}$ is the Fock-space dimension. For a dimer $\mathcal{N} = N{+}1$, while for a trimer
\beq
\mathcal{N}  \ \ = \ \ \frac{1}{2} (N+1)(N+2)
\eeq
The standard representations 
$\Psi_{\bm{n}}=\Braket{\bm{n}}{\Psi}$ is either in the site basis
$\bm{n} = \{ n_j \}$ with $j=1,\cdots,L$, 
or in the momentum-orbital basis $\bm{n} = \{ n_k \}$.  
In this section we list a minimal set of measures that are later used in order to numerically characterize the eigenstates.  

\sect{Quasi momentum}
In the presence of interaction the $n_k$ are not good quantum numbers, but still, without fear of confusion, we can use the notation 
\beq
n_k \ \ \equiv \ \ \braket{\hat{n}_k}
\eeq
Still, due to translation symmetry, the manybody Bloch quasi-momentum $q=(2\pi/L)\times \text{integer}$ remains a good quantum number, and therefore the eigenstates divide into $L$ symmetry classes. For each eigenstate we can calculate the total momentum $P = \sum_k n_k k$. But $P$ is mathematically ill defined and depends on arbitrary modulo convention, whereas the quasi-momentum, unlike $P$, is rigorously a good quantum number:
\beq
q \ \ = \ \ P \ \ = \ \ \sum_k \braket{\hat{n}_k} k, \ \ \ \  \mod(2\pi)
\eeq
The current ${I = (1/L) \sum_k  \braket{n_k} v_k}$ is an optional quantity that can be calculate, where $v_k$ is the velocity of a particle that is placed in the $k$ orbital, see \cite{sfc}. But $I$ unlike $q$ is not a good quantum number.

\sect{Condensation measures}
In the absence of interaction the BHM ground-state is a coherent state where all the particles are condensed in the zero-momentum orbital. As the interaction is increased, the ground-state get-squeezed. We use the following measures to characterize the condensation:  
\beq
n_0 \ &=& \ \braket{\hat{n}_0} \\
\sigma_{0}^2 \ &=& \ \text{Var}(\hat{n}_0)
\eeq
As explained below, we regard $\sigma_{0}^2$ as the measure for the {\em amplitude} fluctuations of the order parameter. 

\sect{Order parameter}
We define ${S_{i,j} \equiv a_i^{\dag} a_j}$. 
For ${i \neq j}$ we use the optional notation 
${S_{i,j} \equiv S_{x}+i S_{y}}$. 
The expectation values $\braket{S_{i,j}}$ are the components of a generalized Bloch vector.  The diagonal elements $\braket{S_{j,j}}$ are the average site occupations. The off-diagonal elements of $\braket{S_{i,j}}$ provide an indication for off-diagonal long range order (ODLRO). 
Due to the symmetry of the system under translations, it depends on the distance ${r=|i-j|}$ mod($L$), and therefore the matrix $\braket{S_{i,j}}$ becomes diagonal once we switch to the momentum-basis. It follows that ODLRO is related to the average occupations $n_k$ of the momentum orbitals, see \App{aA}. At low energies we can regard $n_0$ as the major ODLRO measure. 

\sect{Purity measure}
The one-body reduced probability matrix is 
${ \rho_{i,j} = (1/N)\braket{S_{i,j}} }$.
The purity is defined as
\beq
\mathcal{S} \ = \ \text{trace}(\rho^2)
\eeq
Roughly speaking $1/\mathcal{S}$ is the minimal number of orbitals that are required in order to accommodate the particles. Purity that equals unity defines the notion of coherent state. In the BHM context, coherent state means condensation of all the particles in a single orbital. 
In the vicinity of the SF ground state, the deviation of  $\mathcal{S}$ from unity reflects the depletion ${n_{dep}=N-n_0}$, see \App{aB}. At the SF-to-MI transition the squeezed ground-state turns into a fragmented MI state, that has the lowest purity ($\mathcal{S}=1/L$). 

The purity can also be used to detect self-trapping in a single site. Strictly speaking, due to translation symmetry such states always form ``cat state" superpositions. But in practice any weak disorder breaks the translation symmetry, hence self-trapped states are formed, instead of very narrow solitonic bands.  Such unavoidable symmetry breaking occur also due to numerical noise, or intentionally by introducing weak on-site potential, aka detuning.

\sect{On-site fluctuations}
We use the notation $\hat{n}_{site}$ to indicate any of the $\hat{n}_j$ operators. Due to translation symmetry we have 
$\braket{\hat{n}_{site}}=(N/L)$.  
We define 
\beq \label{eSpar}
\sigma_{\parallel}^2 \ \equiv \ \text{Var}(\hat{n}_{site}) 
\eeq
Later we also define the correlator  
${C(r) = C_{i,j} = \braket{\hat{n}_i \hat{n}_j} }$. 
were $r$ is the ``distance" between the sites.

The entanglement between a given site and the other sites provides an indication for the departure from the regime where GMFT applies. The definition is as follows. We use the standard basis (site occupation). Given an eigenstate $\Psi$ we define   
\beq
\Psi_{n_1,n_2,n_3}  \ &\equiv& \ \Braket{n_1,n_2,n_3}{\Psi} \\
\rho^{(site)}_{n,n'} \ &=& \ \sum_{n_2,n_3}  \Psi_{n,n_2,n_3} \Psi_{n',n_2,n_3}^* \\
\mathcal{S}_{\text{ent}} \ &=& \ \trc[(\rho^{(site)})^2]
\eeq
The matrix $\rho^{(site)}_{n,n'}$ is diagonal, because in each term of the sum we must have ${n=n'=N-n_2-n_3}$ in order to get a non-zero result. 
The diagonal elements are the probabilities $\rho^{(site)}_{n,n}$. Hence $1/\mathcal{S}_{\text{ent}}$ is just another version of the on-site dispersion $\sigma_{\parallel}$.

\sect{Fluctuations of the order parameter}
We already defined $\sigma_{\parallel}$ for the characterization of the  of the single-site fluctuations,  and $\sigma_{0}$ for the fluctuations of the zero-orbital occupation. The latter can be regarded 
as a measure for the fluctuations in the amplitude 
of the order parameter.  Additionally we can define 
a measure $\sigma_{\varphi}$ for the fluctuations in the phase 
of the order parameter. These three measures  
${(\sigma_{0},\sigma_{\varphi},\sigma_{\parallel})}$ 
correspond to the variances of the of the ${(S_x,S_y,S_z)}$ components of the order parameter. 
The technical details regarding the generalization of the ``dimer" Bloch-vector language will be provided in later sections. What we call total fluctuation of the order parameter, denoted as $\sigma_{\perp}^2$, corresponds to the sum of the variances $\sigma_{0}^2$ (amplitude fluctuations) and $\sigma_{\varphi}^2$ (phase fluctuations).

\sect{Higgs measure}
In \Sec{s7} we derive a sum rule from which we can extract $\sigma_{\perp}^2$ given the average occupations $n_k$, 
and the on-site fluctuations $\sigma_{\parallel}^2$. 
Irrespective of that, we calculate the variance $\sigma_0^2$ that characterizes the {\em amplitude} fluctuations of the order parameter. Then we define the Higgs measure as the ratio, namely, 
\beq \label{eHiggs}
\beta =
\frac{\text{\footnotesize amplitude fluctuations in the depletion}}{\text{\footnotesize total fluctuations of the order parameter}}
\ \equiv \ 
\frac{\sigma_0^2}{\sigma_{\perp}^2}
\ \ \ \ 
\eeq
This measure is very small compared with unity for phase oscillations, and becomes of order unity for amplitude oscillations. Of particular interest is the identification of energy levels where the transition from the SF
to the MI phase is non-monotonic, exhibiting relatively large or relatively small $\beta$ at the transition.

\sect{Ergodicity measures}
The participation number $\mathcal{M}$ tells us how many basis states participate in the superposition that forms an eiegenstate.   
Given a basis $\bm{n}$ it is defined as follows:  
\beq
\mathcal{M} = \left[\sum_{\bm{n}} p_n^2 \right]^{-1}
\eeq
where ${p_n = \left|\Psi_{\bm{n}}\right|^2}$.
An individual eigenstate is possibly not ergodic, and does not accommodate the energetically allowed space. In order to determine the volume of the allowed space, we calculate the averaged $p_n$ within a small energy window, and then calculate the associated participation number which we denote as $\overline{\mathcal{M}}$.  
The ratio $\mathcal{M}/\overline{\mathcal{M}}$ serves as a quantum ergodicity measure. For a fully chaotic system such as billiard one expects it to be somewhat less than unity due to fluctuations.  In practice the value is much smaller indicating lack of ergodicity.   
We calculate both $\mathcal{M}_{\text{sites}}$ in the site basis, and  $\mathcal{M}_{\text{orbitals}}$ in the orbital (momentum) basis, and plot   
\beq
\mathcal{M} \equiv \min\left[\mathcal{M}_{\text{sites}},\mathcal{M}_{\text{orbitals}} \right]
\eeq
We note that for ${u\gg 1}$, where the SF-MI transition takes place,    
${\mathcal{M} =\mathcal{M}_{\text{sites}} }$.

\sect{Chaos measure}
In practice it is difficult to associate with an individual eigenstate a measure that indicates whether it is supported by a chaotic sea or by a quasi-regular island. In fact it has been demonstrated in a recent study \cite{sst} that many of the eigenstates do not obey such dichotomy. Nevertheless, we are going to present an efficient method for identification of ``quantum chaos"  via what we call  tomographic inspection of the spectrum.

\section{The phase diagram of the dimer}
\label{s4}

\begin{figure}
\includegraphics[width=4cm]{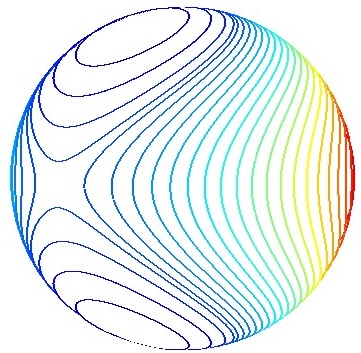}
\caption{{\bf The Bloch Sphere.} 
The phasespace of the dimer system is the two-dimensional Bloch sphere whose embedding coordinates are ${(S_x,S_y,S_z)}$. WKB energy contours ${\mathcal{H}_{\text{dimer}}=E_{\nu}}$ are plotted. For this illustration $u=2.5$ and $N=30$. The sphere is oriented such that the hyperbolic point ${S_x=-N/2}$ is at the front. 
}  
\label{fBloch}  
%
\includegraphics[width=6cm]{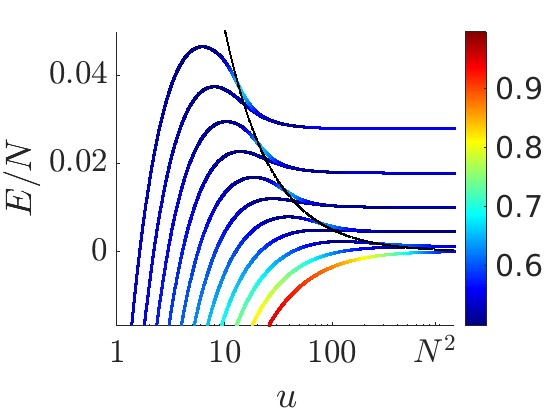}
\caption{{\bf The Energy levels.} 
The lowest energy levels $E_{\nu}$, 
with ${\nu=0,1,\cdots,10}$, versus~$u$. 
The units are chosen such that ${NU=1}$, while $J$ is varied. The levels are color-coded by $\mathcal{S}$. 
The black line is the separatrix energy $E_x$.
}  
\label{dimerEZ}  
\end{figure}

\begin{figure*}

\hspace*{2cm} (a)
\hspace*{4cm} (b)
\hspace*{4cm} (c)
\hspace*{2cm}

\includegraphics[width=5cm]{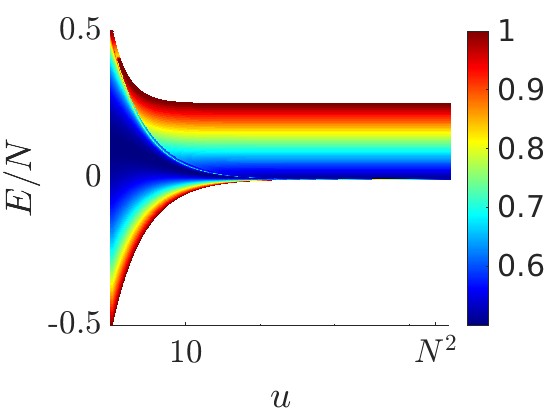} 
\includegraphics[width=5cm]{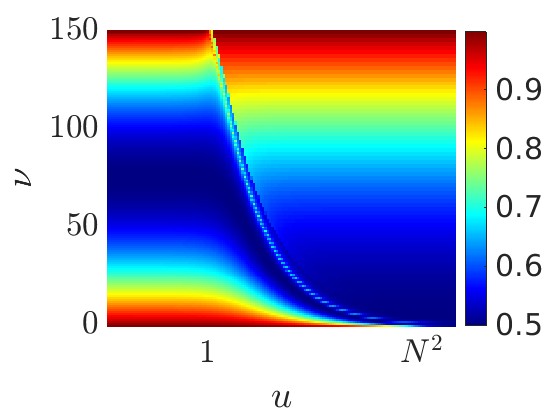} 
\includegraphics[width=5cm]{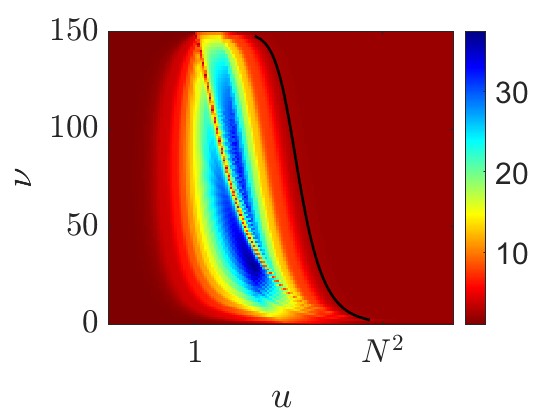}

\hspace*{2cm} (d)
\hspace*{4cm} (e)
\hspace*{4cm} (f)
\hspace*{2cm}

\includegraphics[width=5cm]{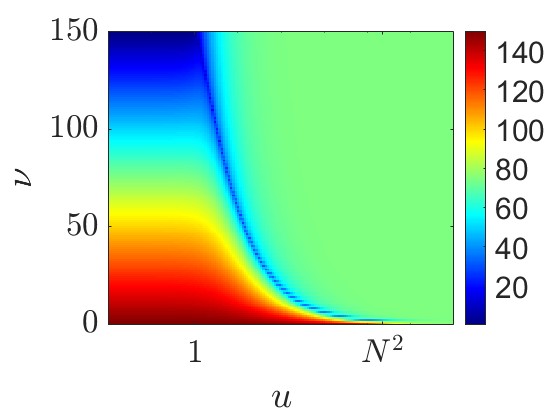} 
\includegraphics[width=5cm]{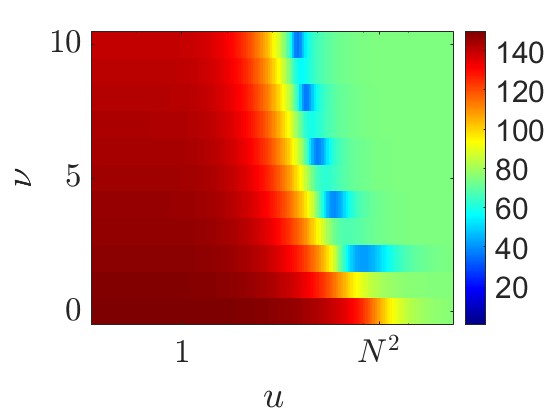}
\includegraphics[width=5cm]{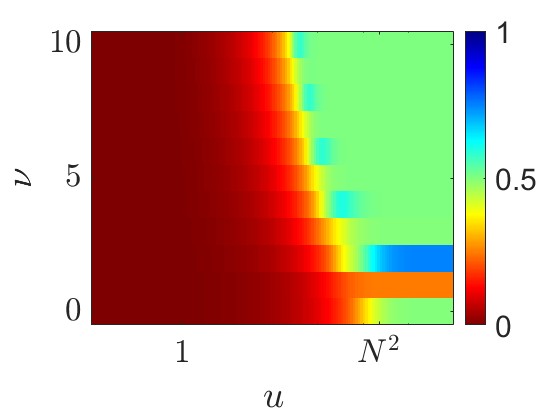}

\caption{
{\bf Phase diagram for the dimer with $N=150$.}
The panels of the first row assume a slightly detuned system in order to identify self-trapping (see text for details). 
(a)~The same as \Fig{dimerEZ}, but including all the levels, hence the lines cannot be resolves. 
(b)~The same plot as in panel~a, as an image, 
where each row represents an energy level $E_{\nu}$, 
ordered with the index $\nu=0,1,2,\cdots$.     
(c)~The same plot as in panel~b, but here the pixels are color-coded by  $\mathcal{M}$. The black line is the border $u_s(E)$ of the perturbative MI regime.  
(d)~Here the color indicates the average occupation $n_0$ of the even orbital.
(e)~The same zoomed. 
(f)~Here the color indicates the Higgs Measure $\beta$. 
Note that green color represents the MI value ${\beta=1/2}$.
}  
\label{dimerE}  
%
\vspace*{3mm}
%
\includegraphics[width=5.5cm]{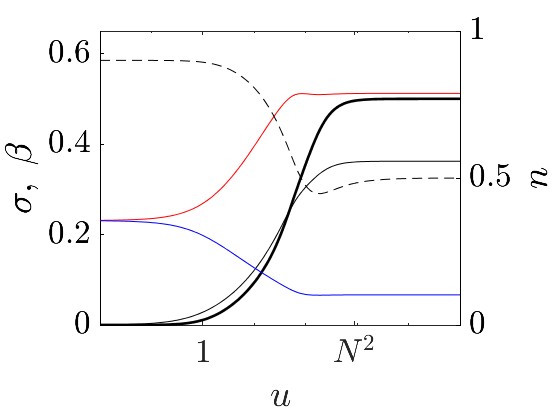}
\hspace*{2cm}
\includegraphics[width=5.5cm]{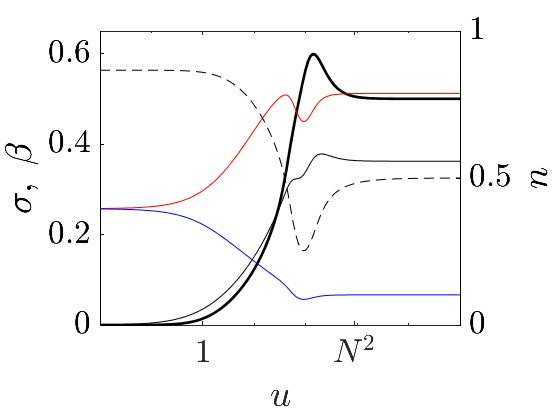}

\caption{{\bf Characterization of fluctuations for dimer eigenstates.} The left and right panels show the variation of the $\sigma$-s for representative odd level ($\nu=3$) and representative even level ($\nu=4$) of an $N=30$ dimer.  
The lines are: 
$\braket{n_0}$ (black dashed, left vertical axis); 
$\sigma_0$ (black);
$\sigma_\perp$ (red);
$\sigma_\parallel$ (blue). 
The Higgs measure $\beta$ (black thick solid line) is the ratio $\sigma_0^2/\sigma_\perp^2$. 
}
\label{HMdimer}  
\vspace*{0mm}
\end{figure*}

Using spin language, and dropping a constant, the dimer Hamiltonian is ${\mathcal{H}_{\text{dimer}} = U S_z^2 - J S_x - \Delta S_z}$. For sake of discussion we have added a detuning potential~$\Delta$ between the two sites. Unless stated otherwise ${\Delta=0}$. 
In the classical limit the dynamics is generated by Hamilton's equations via Poission brackets that assume spin algebra.  
The phase-space of the motion is the Bloch sphere 
${ S_x^2+S_y^2+S_z^2=[(N/2){+}1](N/2) \approx (N/2)^2 }$. 
The classical energy contours 
${\mathcal{H}_{\text{dimer}}(S_x,S_y,S_z)=E}$
feature a seperatrix if ${u>1}$. 
See \Fig{fBloch} for demonstration. 
The energy of the seperatrix equals the energy 
of the unstable hyperbolic point at $S_x=-N/2$, 
that opposes the ground state at $S_x=+N/2$. Namely, 
\beq
E_{min} \ &=& \ -\frac{1}{2}NJ \\
E_x \ &=& \ \frac{1}{2}NJ \\
E_{max} \ & = & \ \frac{1}{4}\left[ u+\frac{1}{u} \right] NJ 
\eeq
We refer to $E>E_x$ as the 
self-trapping (ST) region, while $E<E_x$ 
is the SF region. The latter is diminished and 
cannot accommodate a quantum eigenstates if ${u>N^2}$.   
See \cite{csd} for details and further references.
We plot the eigenenergies $E_{\nu}$ as a function of $u$ in \Fig{dimerEZ} and in \Fig{dimerE}a, and provide further characterization in the additional panels there. Namely, for each eigenstate we calculate the purity $\mathcal{S}$, the ergodicity measure $\mathcal{M}$, the average occupation $n_0$ of the ground-state orbital, and the Higgs measure $\beta$.

\sect{Borders of the MI and SF phases}
If we focus on the ground-state, the definition of the MI phase is unambiguous: it appears for ${u>N^2}$, because eigenstates cannot be accommodated in the SF-region. But if we look at higher energies the notion of MI-phase becomes blurred. There are in fact two relevant borders. The $u_s(E)$ border is determined by perturbation theory (see \App{aE}), and is indicated by black line in \Fig{dimerE}c. As we cross this border from right to left, the MI eigenstates start to mix.
But there is a second border, of the SF phase, as we go from left to right in the diagram. The latter border has to do with the SF separatrix, and is further discussed below.  This border dominates the purity in \Fig{dimerE}b, and the depletion in \Fig{dimerE}d.   

\sect{Spectral perspective}
From a semiclassical perspective the SF transition in the $(u,E)$ diagram is determined by the {\em separatrix}. But in a quantum context one would like to have an unbiased independent determination of the transition.
For a detuned system (${\Delta \neq 0}$), WKB theory implies a {\em minimal} level spacing at the transition, where the classical oscillation frequency $\omega(E)$ vanishes. In the absence of detuning the levels in the MI phase cluster into pairs of quasi-degenerate levels with exponentially small tunnel-splitting.

\sect{Phasespace perspective}
Irrespective of the spectral aspect, the transitions from the ST ro the SF region is characterized by localization at the hyperbolic point. 
We demonstrate this localization by plotting Husimi functions of selected eigenstates. The procedure is defined in \App{aC}, and the plots are displayed and discussed in \App{aD}. 
The observed localization is related to the classical pendulum picture. Namely, at the separatrix energy, where the classical oscillation frequency goes to zero, the dwell time at the ``top" position (the hyperbolic point ${S_x=-(N/2)}$) diverges.      

\sect{ODLRO perspective}
Closer inspection of \Fig{dimerE}e reveals that the localization at the hyperbolic point happens whenever an even superposition crosses the separatrix (the indication for this localization is smaller ${n_0}$). This localization is associated with the appearance of a large Higgs measure $\beta$. A better quantitative inspection over the dependence of $\beta$ on $u$ is provided in \Fig{HMdimer}.

\clearpage 
\section{The phase diagram of the trimer}
\label{s5}

\begin{figure*}

\hspace*{2cm} (a)
\hspace*{4cm} (b)
\hspace*{4cm} (c)
\hspace*{2cm}

\includegraphics[width=5cm]{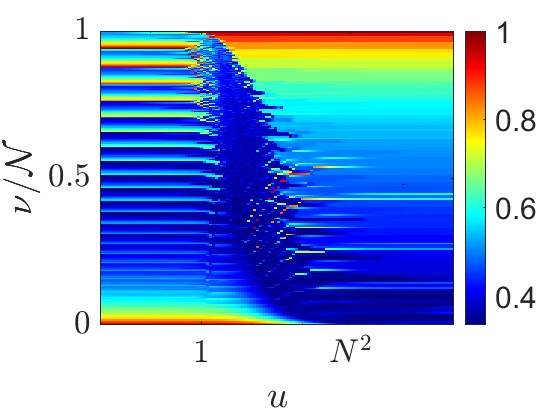}
\includegraphics[width=5cm]{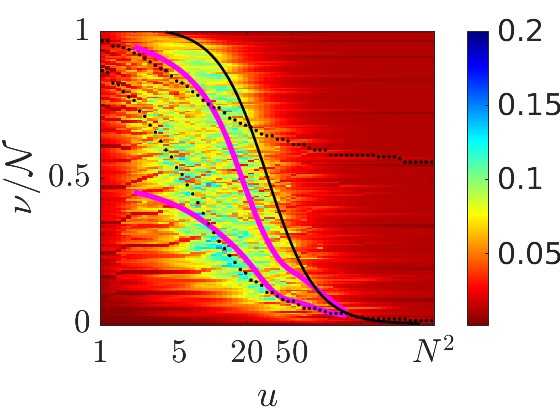}
\includegraphics[width=5cm]{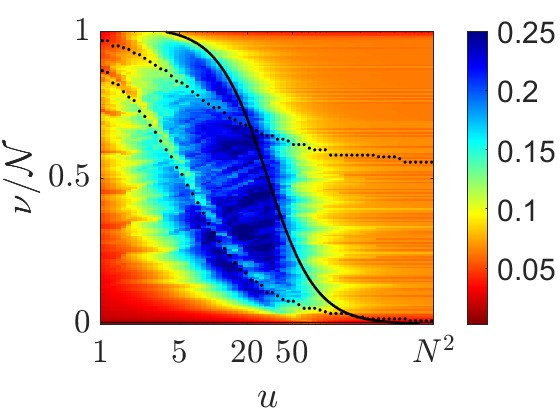}

\hspace*{2cm} (d)
\hspace*{4cm} (e)
\hspace*{4cm} (f)
\hspace*{2cm}

\includegraphics[width=5cm]{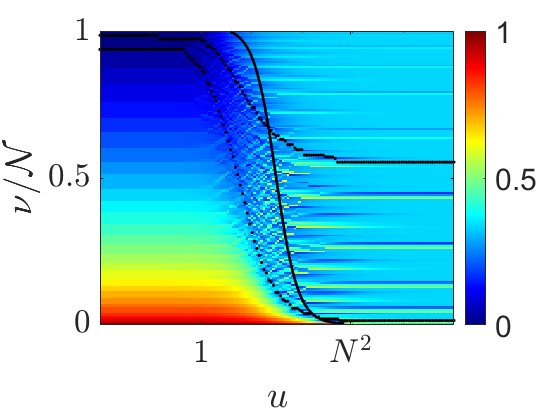}
\includegraphics[width=5cm]{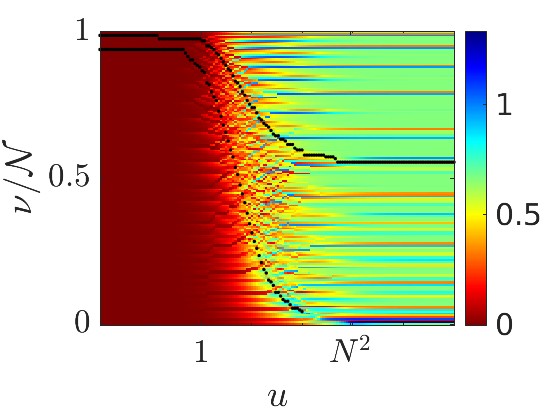}
\includegraphics[width=5cm]{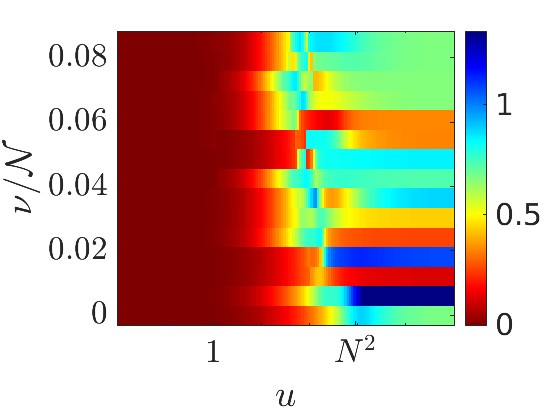}

\caption{
{\bf Phase diagram for the trimer with $N=30$.} 
This is analogous to the dimer diagram \Fig{dimerE}.  
The first panel assumes a slightly detuned system in order to identify self-trapping. The other panels display only the ${q=0}$ states.  
(a)~Image of the Purity $\mathcal{S}$. 
(b)~Image of participation number $\mathcal{M}/\mathcal{N}$.  The dotted black lines indicate the borders $E_{SF}$ (lower) and $E_{ST}$ (higher), and the solid black curve is  $u_s(E)$. The magenta curves indicate the chaos borders.  
(c)~Image of the ergodized participation number $\overline{\mathcal{M}}/\mathcal{N}$. 
(d)~Image of the occupation $n_0$. 
(e)~The Higgs Measure $\beta$. 
Note that green color indicates the MI value ${\beta=2/3}$.
(f)~Zoomed version of the Higgs measure. If we plotted the $q \ne 0$ states, the second level would not exhibit such outstanding large value.    
}
\label{trimerTomog}  
\end{figure*}

\begin{figure*}

\hspace*{2cm} (u=5)
\hspace*{4cm} (u=20)
\hspace*{4cm} (u=50)
\hspace*{2cm}

\begin{minipage}{0.3\hsize}
\includegraphics[width=5cm]{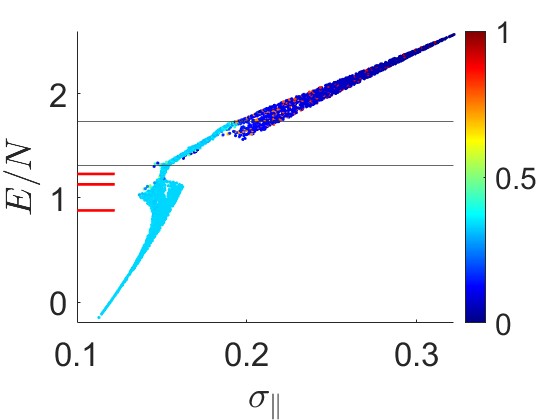} \\
\includegraphics[width=5cm]{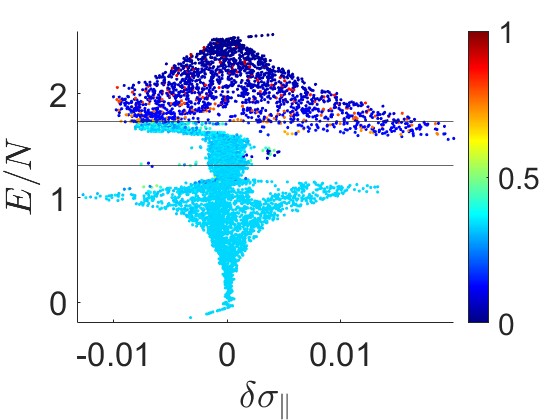} \\
\includegraphics[width=5cm]{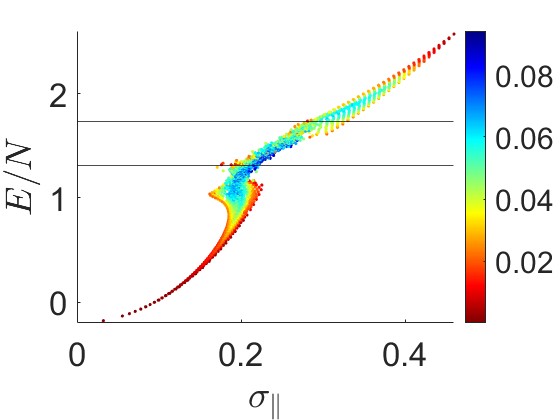} \\
\includegraphics[width=5cm]{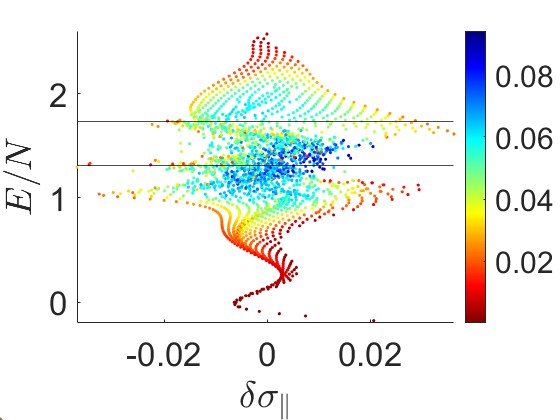}
\end{minipage}
\begin{minipage}{0.3\hsize}
\includegraphics[width=5cm]{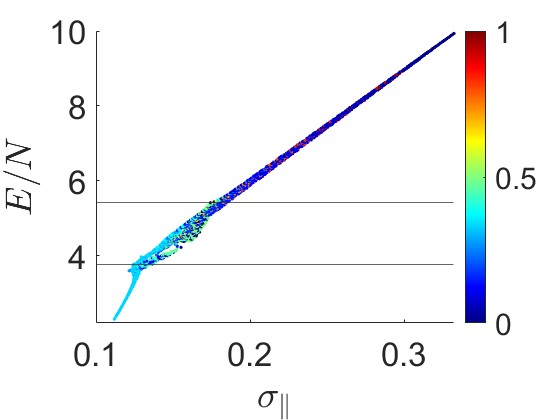} \\
\includegraphics[width=5cm]{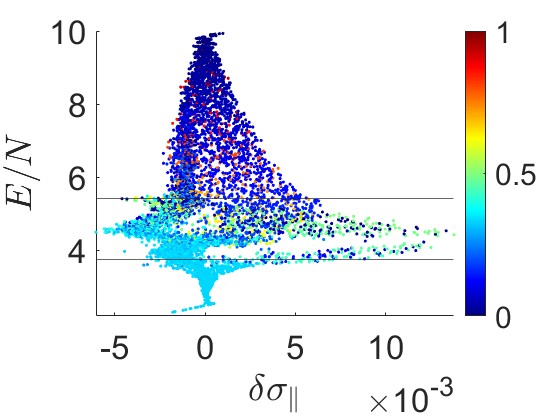} \\
\includegraphics[width=5cm]{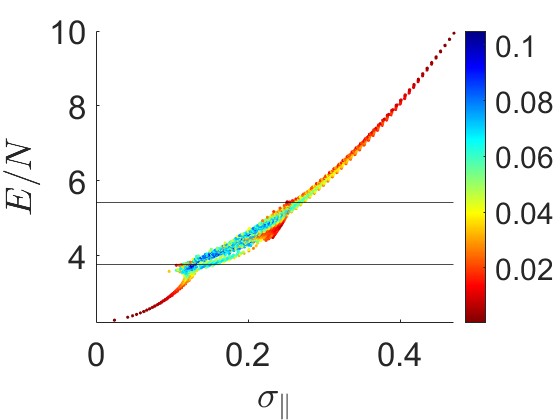} \\
\includegraphics[width=5cm]{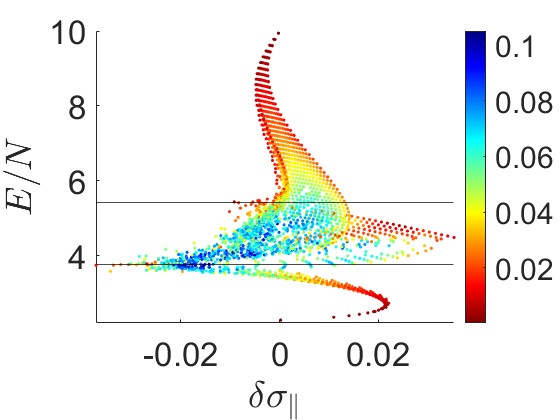}
\end{minipage}
\begin{minipage}{0.3\hsize}
\includegraphics[width=5cm]{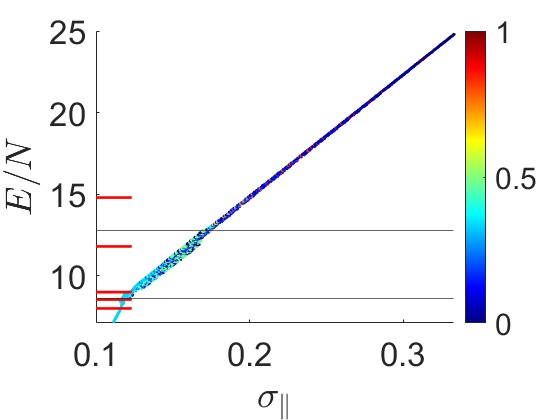} \\
\includegraphics[width=5cm]{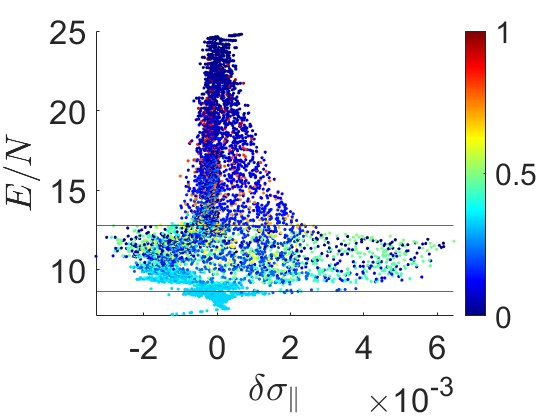} \\
\includegraphics[width=5cm]{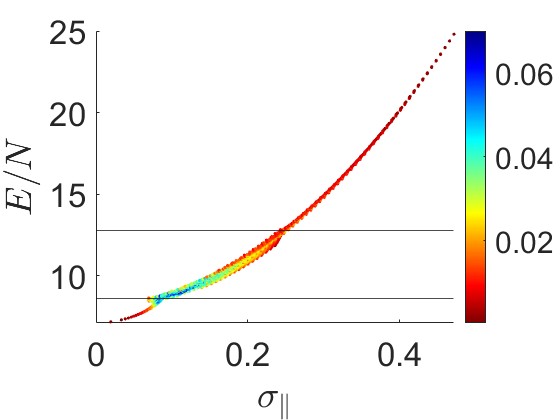} \\
\includegraphics[width=5cm]{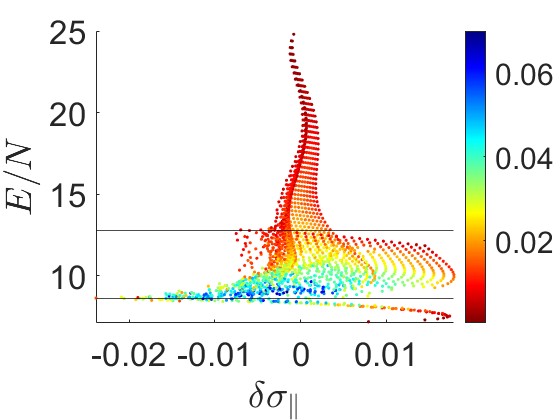} 
\end{minipage}

\caption{
{\bf Trimer spectrum tomography for $N=120$.} 
The columns are for ${u=5, 20, 50}$. 
Rows \#1 and \#2 display the classical spectrum and its zoom. Each point represents a trajectory, and is color-coded by $\bar{n}:= \overline{n_1(t)/N}$, such that blue means ${\bar{n} =0}$ and red means ${\bar{n} =1}$. It is positioned according to its energy~$E$ and its $\sigma_{\parallel}$. 
Rows \#3 and \#4 display the ${q=0}$ quantum spectrum and its zoom. 
Each point represents an eigenvalue, 
and is color-coded by its $\mathcal{M}$.  
It is positioned according to its energy $E_{\nu}$
and its $\sigma_{\parallel}$.   
The units of energy are chosen such that ${J=1}$. 
The horizontal axis is the spreading radius, 
from which the microcanonical average is subtracted
in the zoomed panels.  The horizontal lines indicate the borders $E_{SF}$ and $E_{ST}$.
{\em Technical remark:} In the zoomed version we do not subtract the strict microcanonical average, but a low-order polynomial approximation. Else the patterns in the spectrum are corrupted due to a sampling aliasing. This implies wiggles as an artifact.   
}

\label{trimerSpect}  
\end{figure*}

For each value of $u$ we diagonalize the exact BHM Hamiltonian; find the eigenergies $E_{\nu}$, and calculate for each eigenstate the Purity $\mathcal{S}$ and other measures.   
The global ${(u,E)}$ phase-diagram for the trimer \Fig{trimerTomog}  is obtained by plotting the spectrum for a wide range of $u$ values. {\em Each pixel represents an eigenstate}, and is color-coded by its Purity $\mathcal{S}$ (panel a) or by the ergodicity measures $\mathcal{M}$ (panels b and c), or by the order parameter $n_0$ (panel d), or by the Higgs measure $\beta$ (panels e and f).  
The extracted measures help us to classify the eigenstates. In \Sec{s6} we will take a closer look at representative spectra, for selected values of $u$, that are displayed in \Fig{trimerSpect}. Some representative eigenstates are presented in \Fig{fEig}.

\subsection{Energy landscape}

In the dimer case, the classification of the eigenstates was rather simple, because the structure of the underlying phase-space was determined by the appearance of a single separatrix. In the trimer case, the energy landscape is much more complicated. 
Recall that our standard Fock basis consists of all the possible configurations ${\bm{n}=(n_1,n_2,n_3)}$, with ${n_3=N{-}n_1{-}n_2}$. Thus we get a two-dimensional triangular ${\bm{n}}$-space. 
We refer to it as position space. At each point in this triangular, we define $V_{-}(\bm{n})$ as the floor (minimum), and $V_{+}(\bm{n})$ as the ceiling (maximum) of the energy landscape.  Dropping a constant, the potential floor is 
\beq 
V_{-}(\bm{n}) \ = \  
\frac{U}{2}\sum_i n_i^2  - J \sum_{ \braket{i,j} } \sqrt{n_i n_j}
\eeq
This potential energy is lowest at the center and larger at the corners.  
Along the edges of the triangular region, say along the ${n_3=0}$ edge, the potential surfaces 
are the same as for a dimer:
\beq
V_{\pm} =  \frac{U}{4}\left(N^2 + \Delta n^2\right) \pm \frac{J}{2}\sqrt{N^2-\Delta n^2}   
\eeq
where $\Delta n  = (n_1-n_2)$.
From the above expressions it follows that 
\beq \nonumber
E_{min} &=& \frac{1}{6}N^2U - NJ 
\nonumber \\ 
E_{SF} &=& \frac{1}{6}N^2U + \frac{1}{2} NJ 
\nonumber \\ 
E_{ST} \ &=& \ \frac{1}{4}N^2U + \frac{1}{2} NJ 
\nonumber \\
E_{max} \  &\sim&  \frac{1}{2}N^2U 
\label{eBorders}
\eeq 
%
%
The expression for $E_{SF}$ is determined by $V_{+}$ at the central point of the triangle. For  ${E>E_{SF}}$ the allowed region in $\bm{n}$ space does not contain the center, meaning that the wavefuntion has vanishingly small probability for equal population of the sites. 
The expression for $E_{ST}$, where the allowed region get fragmented, is determined by $V_{+}$ at the mid points of the edges. For $E>E_{ST}$ the wavefunction is concentrated in the 3 corners of the triangular $\bm{n}$ space.

The dimer had a single seperatrix $E_x$ that divides its phase-space into lower SF region and upper ST region.  The trimer has two major separatrixes: the SF region is located at ${E<E_{SF}}$ and the ST region is located at ${E>E_{ST}}$. These borders are indicated in the phase-diagram \Fig{trimerTomog}. It is important to realize that these borders are determined by the energy landscape topography, and do not indicate whether the dynamics is chaotic or not.

\subsection{Non-perturbative mixing}

The left-most SF region and the right-most MI region of the phase-diagram are trivial. These are regions that can be understood within the framework of perturbation theory, either in the {\em orbital} Fock basis or in the {\em site} Fock basis, respectively. 
The site-basis perturbative border is discussed in \App{aE}, leading to the estimate
\beq \label{eUs}
u_{s}(E) \ \ \approx \ \ N\sqrt{\frac{E_{max}-E}{E-E_{min}}} 
\eeq
This border is indicated in \Fig{trimerTomog}b by black line, and will be further discussed below. Its low energy {\em termination} is at ${u_{s} = u_c \sim N^2}$, where the SF-MI transition of the ground-state takes place.

Small $\mathcal{M}$ in \Fig{trimerTomog}b is the indication for the perturbative regions. The measure $\mathcal{M}$ reflects the number of Fock-configurations that are mixed by the Hamiltonian, see \App{aF} for further details. 
In the MI phase of the trimer, the typical value is ${\mathcal{M}\sim 6}$ due to degeneracy that is implied by permutation symmetry.        
As we go in the phase-diagram from right to left, an increase in the value of ${\mathcal{M}}$ indicates level mixing, and eventually provides a circumstantial indication for ``quantum chaos", which involves superposition of many Fock-states.  However, this is not a sufficient condition for its emergence.   

In \Sec{s6} we clarify that the ``chaotic region" features mixed phase-space, meaning that there are sub-regions that are quasi-regular as well as chaotic sea.  This observation is relevant for the characterization and the classification of the eigenstates. We are going to define chaos borders $E_{ch}$ within which most eigenstates are ``chaotic" due to the presence of an underlying chaotic sea. Outside of those borders there are strong fingerprints of ``mixed" quasi-regular regions that support quasi-regular or hybrid eigenstates \cite{sst}.

\subsection{Quantum ergodicity measure} 

The classical accessible area $\mathcal{M}_s$ of the energetically allowed region in the triangular $\bm{n}$-space, is calculated as a function of $E$. See \Fig{trimerPR} for representative plots (upper black line in each panel). In  \App{aF} we provide an analytic estimate and demonstrate that there is reasonable quantum-to-classical correspondence, namely, ${\overline{\mathcal{M}} \sim \mathcal{M}_s}$. 
Coming back to \Fig{trimerPR} we plot in the same panels both $\overline{\mathcal{M}}$ and $\mathcal{M}$ (upper and lower magenta lines respectively). The ratio $\mathcal{M}/\overline{\mathcal{M}}$ can serve as a quantum ergodicity measure.  
Optionally, the quantum ergodicity of the eigenstates can be qualitatively inspected by comparing panels~(b) and~(c) of the phase-diagram in \Fig{trimerTomog}.


\begin{figure}
\includegraphics[width=9cm]{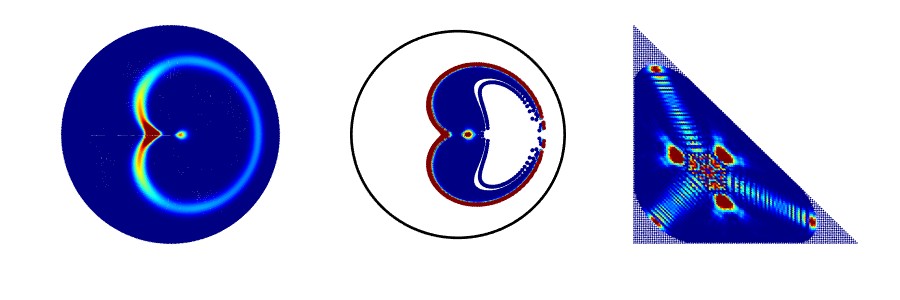}
\includegraphics[width=9cm]{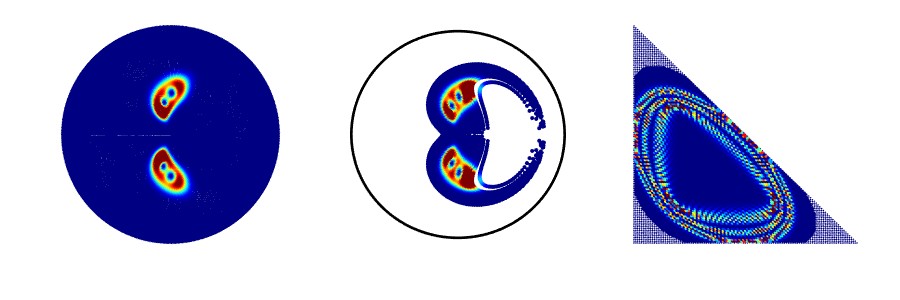}
\includegraphics[width=9cm]{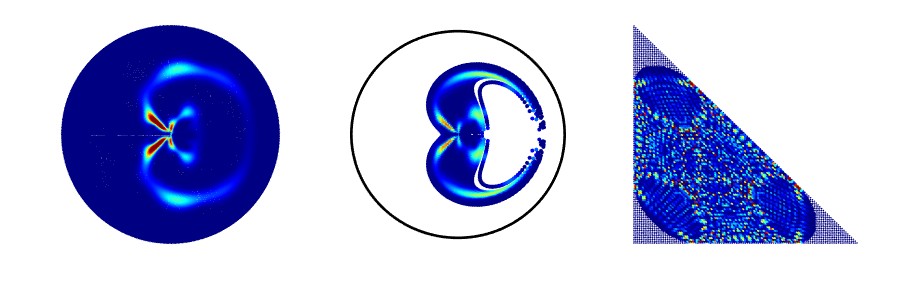}
\includegraphics[width=9cm]{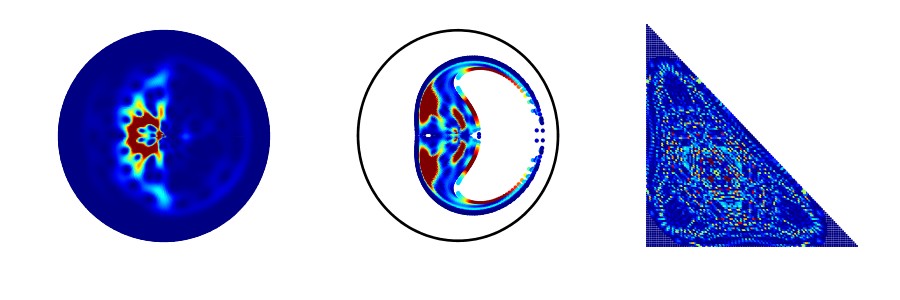}
\includegraphics[width=9cm]{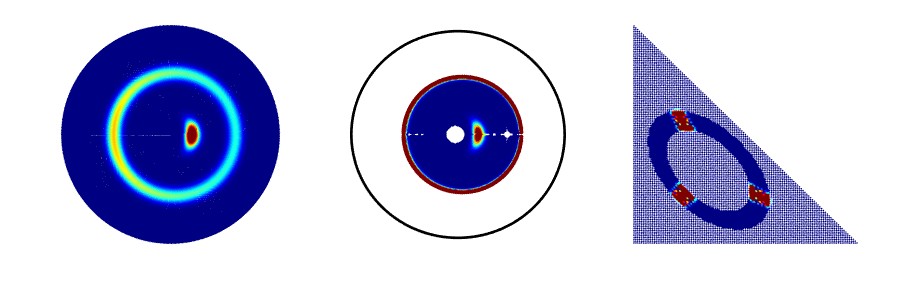}
\includegraphics[width=9cm]{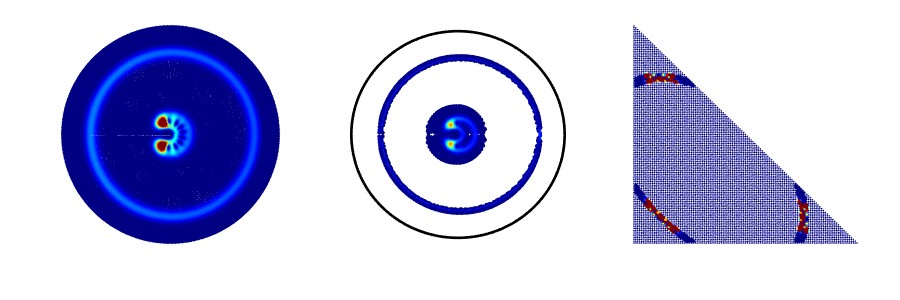}

\caption{
{\bf Representative regular and chaotic eigenstates.} 
Right column: The wavefunction $|\Psi_{n_1,n_2}|^2$. 
Middle column: The Husimi function on the Poincare section.
Left column: Quantum Poincare section.  
The 4 upper panels and the 3 lower panels 
refer to states of the ${u=5}$ and ${u=50}$ spectra of \Fig{trimerSpect}, and are easily associates with regions in the classical Poincare sections of \Fig{trimerPoincare}.  
From top to bottom:
Two regular states at ${E\sim 1}$. 
Chaotic state at ${E\sim 1}$. 
Chaotic state at ${E\sim 1.5}$. 
Regular state at ${E\sim 10}$. 
Self-trapped state at ${E\sim 15}$. 
}

\label{fEig}  
\end{figure}

\begin{figure}
\includegraphics[width=5.7cm]{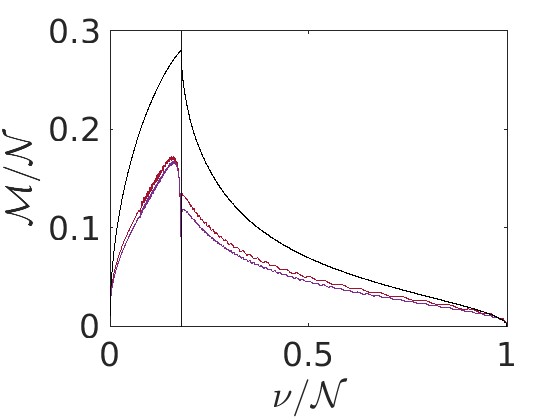} \\
\includegraphics[width=5.7cm]{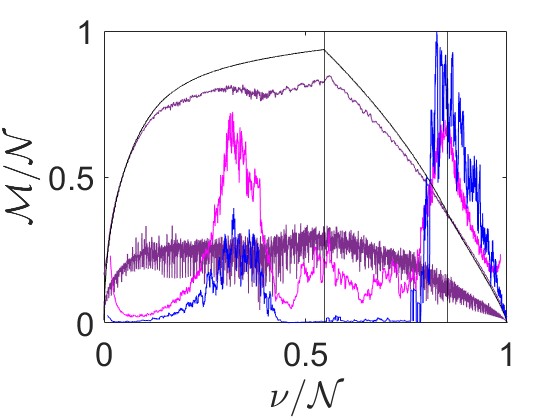} \\
\includegraphics[width=5.7cm]{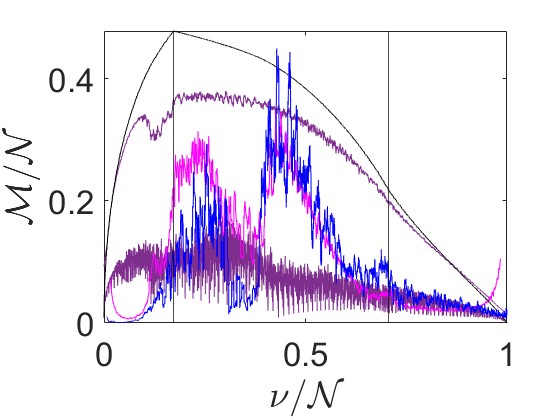} \\
\includegraphics[width=5.7cm]{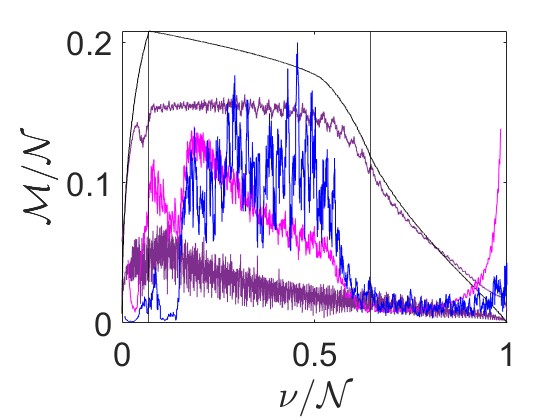} 

\caption{
{\bf The ergodicity measure.} 
The participation number $\mathcal{M}_{orbitals}$ (upper magenta line) of individual eigenstates is plotted versus the scaled energy, and compared with the associated microcanonical  $\overline{\mathcal{M}}_{orbitals}$ (lower magenta line). We also plot (black lines) the classically accessible area~$\mathcal{M}_s$.
Upper panel:~Dimer case ${u=50}$. The curves are for ${N=1000,2000}$ (barely resolved). The vertical line indicates the separatrix $E_x$. 
Other panels:~Trimer case ${u=5,20,50}$. We plot the result for the $q{=}0$ eigenstates. The $q{\neq}0$ plots (not displayed) look the same.  
The curves are for ${N=72,120}$ (barely resolved). 
Vertical lines indicate $E_{SF}$ and $E_{ST}$. 
We also plot $\Delta\sigma_{\parallel}$ that is extracted from the classical and from the quantum spectra (blue and pink lines respectively). 
}  
\label{trimerPR}  
\end{figure}

\subsection{The SF-MI transition} 

It is now appropriate to provide a precise meaning for the notion of ``SF-MI transition". We are not focusing here on the ground-state. We are looking on the full spectrum. While the quantum phase-transition of the ground-state is established in the thermodynamics (large~$L$) limit, the question arises whether some kind of {\em mobility edge} extends to higher energies. We do not have the tools to establish the existence of such rigorous {\em phase-transition} line, but we do have a way to argue that a transition takes place, and to clarify its borders.  

Going in the phase-diagram from left to right, we realize that  there is a quantum signature for the the separatrix $E_{SF}$ that bounds the SF-region. For the dimer the transition is rather sharp, see \Fig{dimerE}. It signifies simple classification of eigenstates into those that reside in the SF-region (below the separatrix) and those that reside in the ST-region (above the separatrix). It is a {\em classical} border that features a low energy {\em quantum} termination at ${u_c \sim N^2}$. Namely, as far as the {\em ground-state} is concerned, the relevant question is whether a squeezed coherent state can be accommodated by the ${E<E_{SF}}$ region. If this region is less than Planck cell, the ground state becomes MI-like. This leads to the identification of the {\em quantum} phase transition at ${u_c\sim N^2}$, and motivates the definition of the quantum parameter $\gamma$ of \Eq{eGamma}. 

The $E_{SF}$ border is apparent also in the trimer diagram \Fig{trimerTomog}, but it is somewhat blurred. The reason for the blurring is the chaos that emerges in the vicinity of the separatrix. This statement is further supported by comparison with the Bogolyubov approximation where chaos is absent, see \App{aG} for details. Whether this border becomes a sharp {\em mobility edge} in the ``thermodynamics limit" (${L\rightarrow \infty}$) is a matter for speculations.  

An optional way to identify the SF-MI transition is to inspect the spectrum of a rotating ring. In the SF-phase we expect sensitivity to $\Phi$. The extreme sensitivity is expressed as symmetry breaking of the ground-state, as in first-order phase transition. In \App{aH} we demonstrate this symmetry breaking by considering the spectrum of a rotating trimer with ${\Phi \sim 3\pi}$, for which the ${q=\pm 2\pi/3}$ condensates are quasi-degenerated. 

We now turn to discuss the MI-to-SF transition as we go from right to left in the phase-diagram. Let us look first at panel~c of the dimer diagram  \Fig{dimerE}.  The black line indicates the perturbative border $u_s(E)$ where levels start to mix. But this border does not signify a phase-transition. The ``deformation" of the eigenstates in phase space is gradual, and does not involve strutural changes. It is somewhat analogous to the squeezing that is observed in the SF side of the transition. 

The perturbative border $u_s(E)$ becomes more interesting for the trimer. Recall that in the dimer case each energy-band consists of two quasi-degenerated states, while for the trimer the typical quasi-degeneracy of each energy-band is 6, due to permutation symmetry. It is illuminating to observe that away from the ground state ${u \sim u_s(E)}$ implies ${\mathcal{M}_s(E) \sim N}$. It means that the couplings between Fock-states of the annulus-shaped allowed-region in $\bm{n}$-space form a {\em percolating cluster}. This allows the formation of ODLRO. Indeed as we cross $u_s(E)$ in the phase-diagram from right to left, fluctuations in the ODLRO appear, which we further discuss in \Sec{s9}. But those fluctuations do not indicate systematic structural changes, as opposed to $E_{SF}$.


\section{Spectrum tomography}
\label{s6}

In practice it is difficult to associate with an individual eigenstate a measure that indicates whether it is supported by a chaotic region. The common practice is to look on the level statistics. But such approach has two issues: 
{\bf (i)}~It provides numerically clear results only for rather simple systems with large $N$, such that the spectrum in the range of interest is dense enough; 
{\bf (ii)}~It does not allow classification of the eigenstates in the typical situation of underlying mixed chaotic and quasi-regular dynamics. We therefore suggest below to adopt a different strategy that we call spectrum tomography. This tomography has both quantum and corresponding classical versions.

\subsection{Classical spectrum tomography} 

Classical ergodicity can be quantified by launching a cloud of trajectories that have a given $E$. For globally chaotic phase space, due to ergodicity, the time-average of a given observable will be the same for all the trajectories, whereas for a mixed phase space we expect a wide distribution. This opens the possibility to easily identify  mixed phase-space regimes, which we further discuss below.
In practice we generate a uniform cloud of trajectories in phase space. Each trajectory is characterized by its energy $E$ and by the dispersion $\sigma_{\parallel}$ of $n_{site}$. Accordingly, each trajectory provides a point ${ (\sigma_{\parallel},E) }$ of what we call classical spectrum, see \Fig{trimerSpect}. 

A technical computational remark is in order. In the quantum spectrum, the average of $n_{j}$ for each eigenstate, is strictly $1/L$. But for e.g. a self-trapped classical trajectory it is not so. The temporal average $\bar{n}$ of the first site (${j=1}$) is calculated for each trajectory, and is indicated by color in the plot of the classical spectrum. This helps to identify classical self-trapping.  In the calculation of the classical dispersion $\sigma_{\parallel}$, that coresponds to \Eq{eSpar}, the temporal average is taken over all the sites (${j=1,2,3}$), such as to guarantee proper correspondence with the associated quantum cat-state superposition (${\bar{n}=1/3}$).

\subsection{Quantum spectrum tomography} 

The quantum spectra of \Fig{trimerSpect} correspond to the classical spectra. Each point ${ (\sigma_{\parallel},E) }$ in the quantum spectrum represents an eigenstate, and is color-coded by its $\mathcal{M}$. It is positioned according to its energy $E_{\nu}$, and its $\sigma_{\parallel}$. The latter can be regarded as an indication for the location of the ``wavefunction" in $\bm{n}$ space. For example: the SF ground state is located at the center of the triangular $\bm{n}$ space, and therefore has a very small $\sigma_{\parallel}$; in contrast the upper states are superpositions of self-trapped configurations that are concentrated at the corners of the triangular $\bm{n}$ space, and therefore have very large $\sigma_{\parallel}$. For eigenstates in the chaotic sea the values of $\sigma_{\parallel}$ bunch randomly around a microcanonical average value, whereas in quasi-regular regions they form a lattice-like arrangement that reflects EBK quantization.  For further discussion of the associated Monodromy see \cite{bhm}.

\sect{Technical efficiency} 
The tomographic quantum spectrum can be regarded as a blurred or coarse-grained version of the classical spectrum. Since it roughly contains the same information, it offers an {\em efficient} way to numerically study classical dynamics: the cost of producing a quantum spectrum via instantaneous diagonalization of a matrix is negligible compared with the cost of producing a multi-trajectory classical spectrum.  Even if Nature were classical, a quantum procedure would be of great value.

\begin{figure*}

\includegraphics[width=5.5cm]{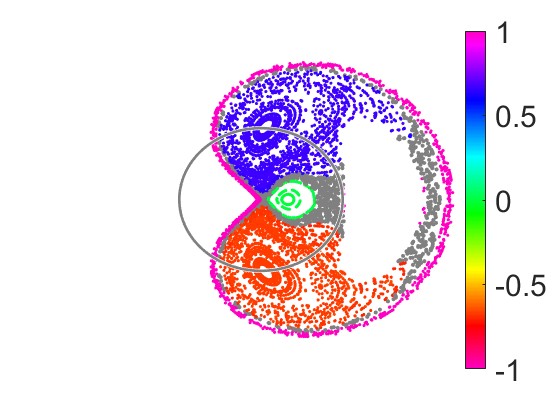} 
\includegraphics[width=5.5cm]{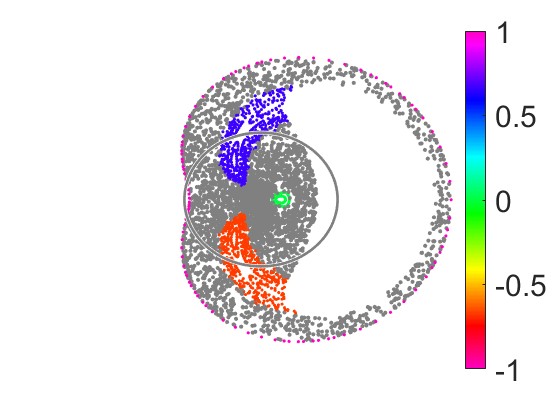}
\includegraphics[width=5.5cm]{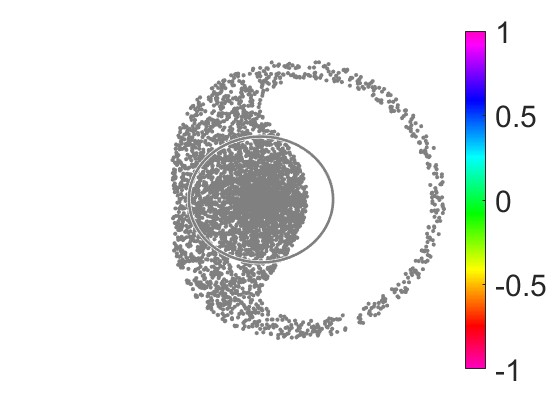}
\\
\includegraphics[width=5.5cm]{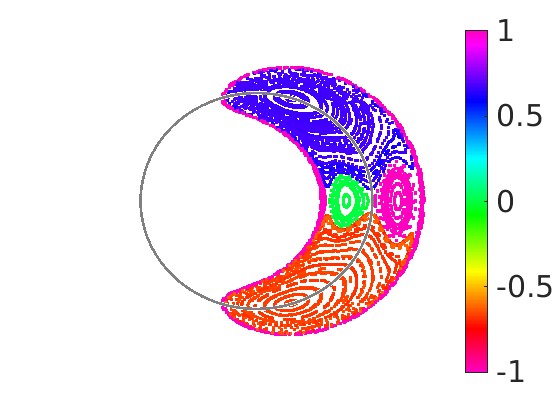} 
\includegraphics[width=5.5cm]{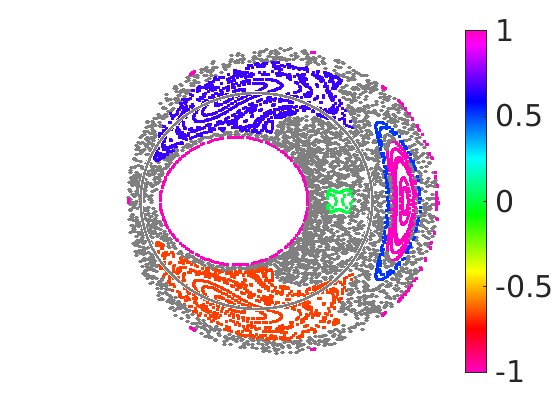}
\includegraphics[width=5.5cm]{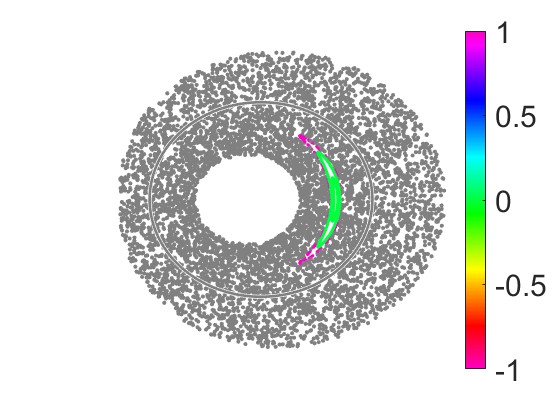}
\\
\includegraphics[width=5.5cm]{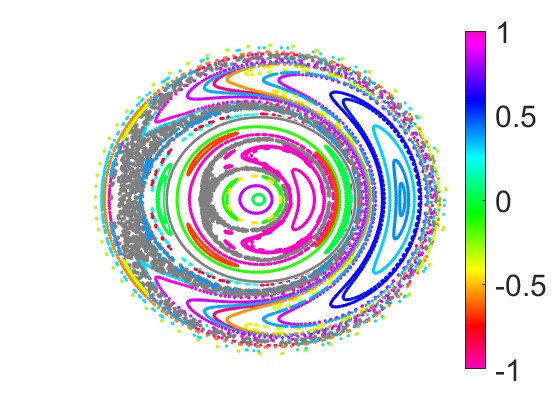}
\includegraphics[width=5.5cm]{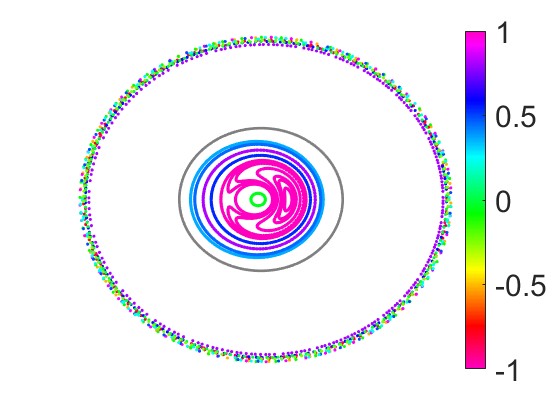}

\caption{
{\bf Representative Poincare sections.} 
The section is defined in the main text. 
The polar coordinates of the panels are $(\varphi_{site},n_{site})$. 
Panels of the first row are for $u=5$, 
while the other panels are for $u=50$. 
The section energies in ascending order from left to right are indicated by small horizontal bars in the upper panels of \Fig{trimerSpect}.  
%
Each quasi-regular trajectory is colored by its~$k/\pi$. 
The extraction of the wavenumber~$k$ is clarified in \Fig{trimerTraj}. If $k$ is numerically ill defined, it indicates chaos, and then the trajectory is colored in gray.
}
\label{trimerPoincare}  
%
\vspace*{1cm}
%
\includegraphics[width=5.5cm]{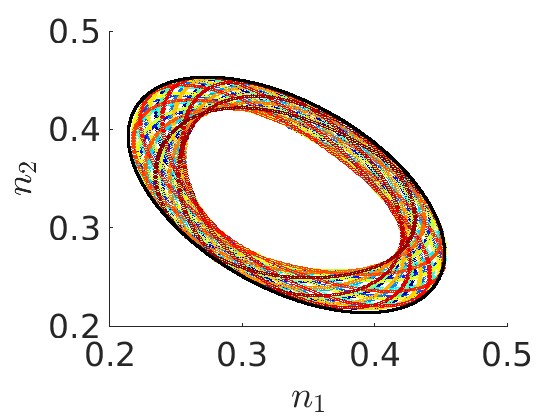}
\includegraphics[width=5.5cm]{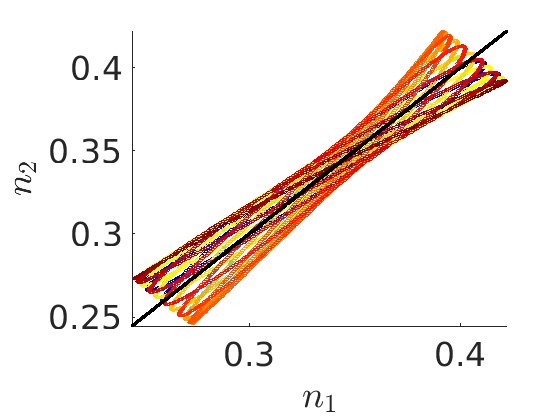} 
\includegraphics[width=5.5cm]{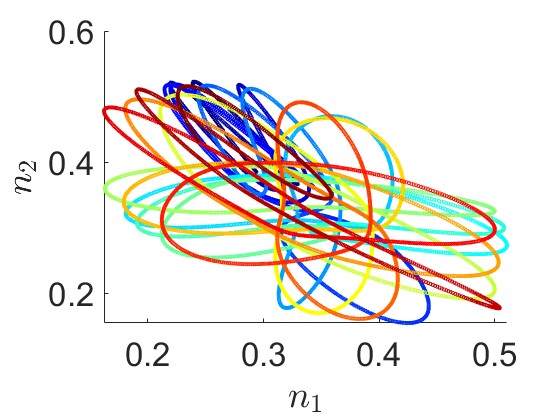} 

\caption{
{\bf Regular and chaotic modes.} 
The major regular classical trajectories are associated with the ${k=0}$, and with the ${k=\pm 2\pi/3}$ Bogulyubov modes. Representative plots of $(n_1(t),n_2(t))$ are displayed, from which $k$ is extracted. The left panel illustrates the ${k=+2\pi/3}$ locking of a $u{=}50$ purple trajectory in \Fig{trimerPoincare}. The ${k=- 2\pi/3}$ version (not displayed) looks the same, but the trajectory is explored in the opposite sense. The middle panel illustrates the ${k=0}$ locking of a green trajectory in the same figure. In the latter case, there are associated plots for the other bonds (not displayed) that exhibit ${k=\pi}$ locking, as implied by conservation of particles.  The right panel shows a chaotic trajectory for which $k$ is ill defined.   
}  
\label{trimerTraj}  
\end{figure*}

\subsection{Underlying phasespace}

We can regard the BHM Hamiltonian as describing a chain of coupled oscillator. The trimer is formally a two degree of freedom system with canonical coordinates $(n_1,n_2)$ and conjugate coordinates 
${\tilde{\varphi}_1 = (\varphi_1-\varphi_3)}$
and ${\tilde{\varphi}_2 = (\varphi_2-\varphi_3)}$. 
The structure of phase-space can be illustrated using Poincare sections at different energies. The section is defined by 
${\tilde{\varphi}_2 = 0}$. 
Some representative Poincare sections that illustrate the dynamics are provided in \Fig{trimerPoincare}. The polar coordinates of the plots are $(\varphi_{site},n_{site})$, where $n_{site} = n_1$ and  ${ \varphi_{site}=\tilde{\varphi}_1 }$.  

For large $u$ and low $E$ the allowed region is around the central stationary point ${\text{CSP}=(\varphi{=}0,n{=}1/3)}$. The dynamics is quasi-regular, reflecting ``small vibrations" of the chain.  Specifically, the Bogulyubov analysis predicts modes that correspond to the wavenumbers ${k=(2\pi/L)\times \text{integer}}$.
Indeed we see that the quasi-regular region is divided into 5 sub-regions. Two regions are centered around secondary fixed-points of the Poincare section that represent small $n_{site}$ oscillations around the CSP with relative phase difference ${k=\pm 2\pi/3}$, as illustrated in the left panel of \Fig{trimerTraj}. The 3 other regions feature ${k=0}$ and $k=\pm\pi$.  
%
%
Note that all the bonds of a given trajectory $(n_1(t),n_2(t),n_3(t))$ either feature ${k=0}$ oscillations, or else two of them feature ${k=\pm \pi}$ oscillations, such that the total number of particles is conserved.  

For higher energies, see \Fig{trimerPoincare}, the seperatrix region becomes chaotic, and turns into a chaotic sea. For much larger $E$ the energy surface become fragmented into small $n_{site}$ disc and large $n_{site}$ annulus, reflecting self-trapping in a single site.

The quasi regular and the chaotic regions in phase space support eigenstates of the BHM Hamiltonian. Representative Husimi functions of the representative eigenstates are provided in \Fig{fEig}.

\subsection{Chaos borders} 

We can inspect the dispersion in the values of $\sigma_{\parallel}$ in order to quantify the {\em chaoticity} of a given energy shell. Let us look for example in the spectrum of the ${u=5}$ trimer that is displayed in \Fig{trimerSpect}. We define a microcanonical average $\overline{\sigma_{\parallel}}$ that is evaluated in each energy bin. Then we look on deviations from this average value, namely,     
${ \delta \sigma_{\parallel} = \sigma_{\parallel} - \overline{\sigma_{\parallel}}  }$, and define the associated dispersion 
\beq
\Delta  \sigma_{\parallel} \ \  = \ \ \sqrt{\overline{(\delta \sigma_{\parallel})^2}}
\eeq
The variation of $\Delta\sigma_{\parallel}$ as a function of $E$ is illustrated in \Fig{trimerPR}. It helps to figure out the range of energy where ``quantum chaos" prevails. 
Coming back to \Fig{trimerSpect}, as we go up in energy $\Delta  \sigma_{\parallel}$ becomes larger, indicating a larger fraction with quasi-regular motion. But there is a rather sharp value ${E=E_{ch}}$ above which $\Delta  \sigma_{\parallel}$ shrinks, indicating that quasi-regular eigenstates, of the type that are displayed in the two upper rows of \Fig{fEig}, have been mixed with neighboring quasi-regular eigenstates, and now become part of a global chaotic sea. An example for a chaotic eigenstate in this `global' chaotic sea is provided in the 3rd line of \Fig{fEig}. Similarly, starting at the top energy, as we go down, we can define an upper chaos threshold ${E=E_{ch'}}$. What we call chaotic range is the energy interval ${ E_{ch}< E < E_{ch'} }$,  see \Fig{trimerTomog}c. This interval shrinks as $u$ is increased, and eventually diminishes.   

The two chaos borders ${E_{ch}}$ and  ${E_{ch'}}$ are determined in practice by inspection of $\sigma_{\parallel}$ plots as in \Fig{trimerPR}. One would expect that they would give indication for the region where $\mathcal{M}$ and $\overline{\mathcal{M}}$ are correlated. In practice we see that the correlation between $\mathcal{M}/\overline{\mathcal{M}}$ and $\sigma_{\parallel}$ is rather weak: quantum ergodicity is lacking also in regions where the underlying classical dynamics is globally chaotic. We conclude that the ``roughness" of the chaotic sea is as effective as the existence of quasi-regular regions. Namely, both lead to phase-space localization of the eigenstates. We believe that this is typical for systems with weak chaos, and/or few degrees of freedom.

\section{Characterization of fluctuations}
\label{s7}

It is pedagogically illuminating to use the semiclassical Bloch-sphere picture of the dimer in order to discuss the characterization of fluctuations. This language can be extended to any $L$ site system as discussed in \App{aA}. In this section, for the purpose of providing a simplified introduction, we assume ${N \gg 1}$, and ignore relative error of order~$1/N$. 

One can regard an eigenstate of the dimer as a cloud of points on the Bloch sphere (technically speaking we can use a Huismi function for visualization). Any point of the cloud satisfies ${S_x^2+S_y^2+S_z^2 \approx (N/2)^2}$, 
with the identification ${S_z \equiv n_{site}-(N/2)}$, 
and ${S_x \equiv n_0-(N/2)}$.  
Averaging over the cloud one deduces that  
${\braket{S_x}^2 + \sigma_x^2 + \sigma_y^2+ \sigma_z^2=(N/2)^2}$, which assumes ${\braket{S_y} = \braket{S_z} =0}$ due to symmetry. 
Given $\langle \hat{n}_0 \rangle$ and 
${\sigma_x^2 = \sigma_0^2 }$
and ${\sigma_z^2 = \sigma_{\parallel}^2 }$ 
we can extract ${\sigma_y^2 \equiv \sigma_{\varphi}^2 }$.  
The relation can be written as  
\beq
\sigma_{\parallel}^2 + \sigma_{\perp}^2 \ \approx \ (N-n_{dep})n_{dep}
\eeq
where ${n_{dep}=N-n_0}$ is the depletion, 
and ${\sigma_{\perp}^2=\sigma_0^2 + \sigma_{\varphi}^2}$ 
is what we call total fluctuations of the order parameter.  
The analogous relation for the trimer is 
\beq \label{eSRSF}
\sigma_{\parallel}^2 + 2\sigma_{\perp}^2 
\ \approx \ \frac{2}{3}N n_{dep}
\eeq
which assumes small depletion (${n_{dep} \ll N}$). 
A precise version of these relations, and their generalization for an $L$ site ring, are discussed in the following subsections. They are useful for the calculation of the Higgs measure $\beta$, and for getting insights.  

\begin{figure}
\includegraphics[width=8cm]{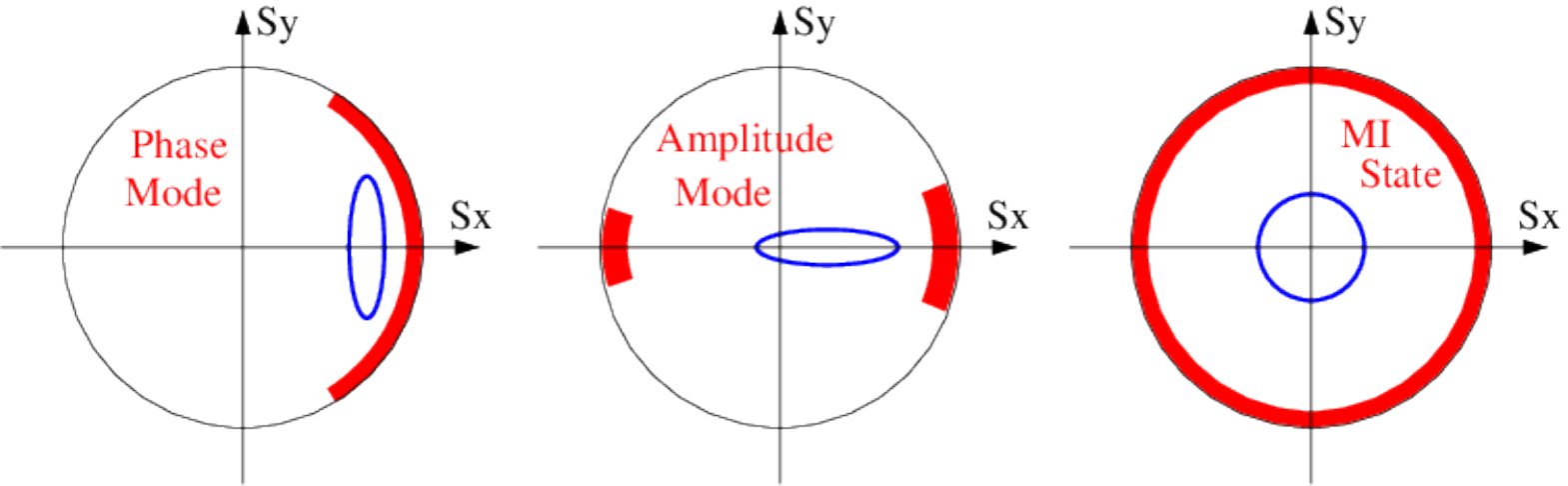}
\caption{ {\bf Visualization of the $\sigma$ measures.}
Caricature of Husimi distributions (red clouds) on the Bloch sphere, projected onto ${(S_x,S_y)}$.  From left to right: Phase-mode eigenstate; 
Amplitude-mode eigenstate; and Mott-Insulator ground state. The 3 states feature ${ S_z \sim 0 }$, meaning that the particles are roughly divided equaqlly between the two sites (with uncertainty that diminishes in the MI case).   
The red color indicates the phase-space region where the cloud spreads, 
which is along the equator. The blue line indicates an 
ellipse that is centered at ${(\braket{S_x},\braket{S_y})}$ 
with major axes ${(\sigma_x,\sigma_y)}$. The ratio of the major axes determines the amplitude/phase character of the eigenstate.  
}
\label{fC}  
\end{figure}

For visualization purpose we note that 
any eigenstate can be represented by 
an ellipsoid whose major axes are  
${(\sigma_{0},\sigma_{\varphi},\sigma_{\parallel})}$. 
The actual phase-space distribution might look very different. 
This statement is clarified using the caricature of \Fig{fC}.

\subsection{The ODLRO fluctuations}

Given the average occupations $n_k \equiv \braket{\hat{n}_k}$,  we can calculate the components $\braket{S_{i,j}}$ of the generalized Bloch vector. Assuming clean ring with translation symmetry, we always have 
\beq
\braket{S_{j,j}} \ \ = \ \ \braket{\hat{n}_j} \ \ = \ \ \frac{N}{L}  
\eeq
More generally we have the relation  
\beq \label{eSr}
S(r) = \braket{S_{i,j}} 
= \braket{a_i^{\dag} a_j} 
\ = \ \frac{1}{L} \sum_k n_k \, e^{-ikr} 
\eeq
Here ${r=(i-j)}$ is a dummy index that reflects the ``distance" mod($L$) between the sites. Full condensation in the zero momentum orbital implies full ODLRO with ${S(r) = N/L}$. Useful expressions for the $S(r)$ of the dimer and the trimer are provided in \App{aA}.  The fluctuations of the site occupations are characterized by 
\beq
C_{i,j} \ = \ \braket{\hat{n}_i \hat{n}_j} \ \equiv \ C(r)
\eeq
The diagonal elements are ${C(0) \equiv (N/L)^2 + \sigma_{\parallel}^2}$. 
The correlation between occupations of different sites is related to the fluctuations $\sigma_{\perp}^2$ of the order parameter, through the relation: 
\beq \label{eDefF}
C(r) = \braket{a_i^{\dagger}a_i a_j^{\dagger}a_j} 
\ \equiv \ |S(r)|^2  - \frac{N}{L} + \sigma_{\perp}(r)^2 
\ \ \ \ 
\eeq

\subsection{The ODLRO sum rule}

A sum rule over the fluctuations is implied from conservation of particles. Squaring the sum ${\sum_j \hat{n}_j = N}$, and taking the expectation value we deduce that 
\beq
\sigma_{\parallel}^2 + \sum_{r=1}^{L{-}1} C(r) 
\ = \ (L{-}1)\left(\frac{N}{L}\right)^2
\eeq
Using \Eq{eDefF}, this can be written as 
\beq \label{eSR}
\sigma_{\parallel}^2 + 
\sum_{r=1}^{L{-}1} \left[ |\vec{S}(r)|^2 + \sigma_{\perp}(r)^2  \right]
= (L{-}1) \frac{N}{L} \left( \frac{N}{L} {+} 1 \right) \ \ \ \ 
\eeq
For the dimer and for the trimer the indication for dependence on $r$ (mod($L$)) can be omitted because we have only one ``distance", namely, ${|r|=1}$.        
For the dimer it follows that
\beq \label{eSRd}
\sigma_{\parallel}^2 + \sigma_{\perp}^2 
\ = \ \frac{N}{2} \left( \frac{N}{2}+1 \right) 
- \left(n_0-\frac{N}{2}\right)^2 
\eeq
For the trimer it follows that
\beq \label{eSRt}
\sigma_{\parallel}^2 + 2\sigma_{\perp}^2 
\ &=& \ 2\frac{N}{3} \left( \frac{N}{3}{+}1 \right)
- \frac{1}{2}\left(n_0{-}\frac{N}{3}\right)^2 
\nonumber \\
&& - \frac{1}{6}(n_{+}{-}n_{-})^2
\eeq
%
%
%
Thus we can deduce the total fluctuations $\sigma_{\perp}^2$ of the order parameter, after subtraction of the on-site fluctuations.

\section{The Higgs measure}
\label{s9}

\begin{figure}
\includegraphics[width=6cm]{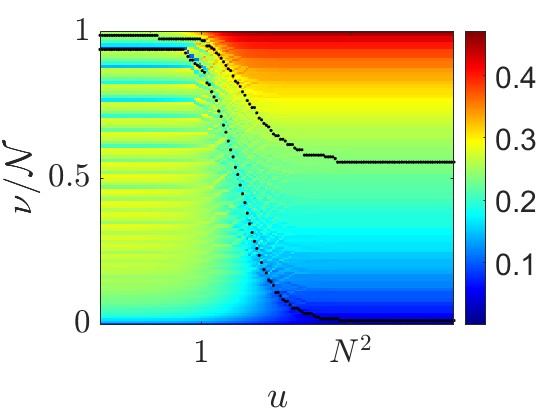} 
\includegraphics[width=6cm]{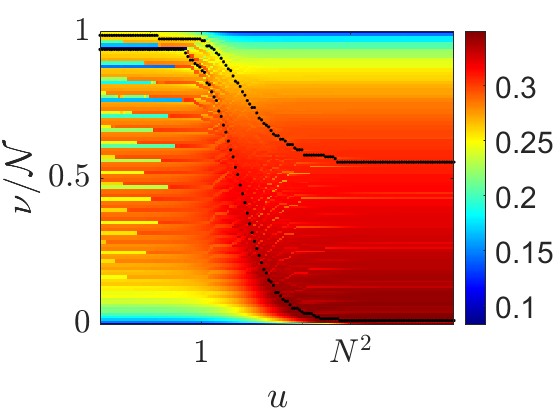}
\includegraphics[width=6cm]{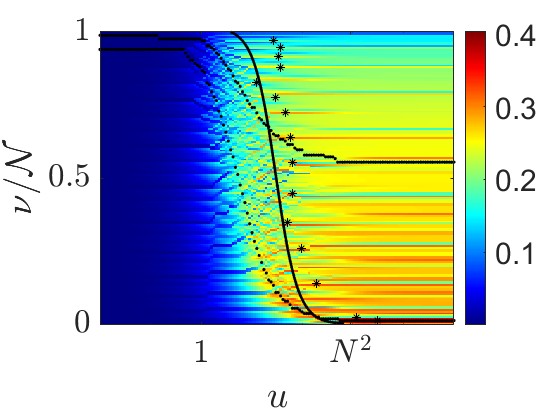}
\caption{
{\bf Fluctuations phase diagram.} 
Images of $\sigma_{\parallel}$, and $\sigma_{\perp}$,  and $\sigma_0$ for the $q=0$ eigenstates of the phase-diagram. Namely, these are the raw input data for the calculation of~$\beta$ in \Fig{trimerTomog}e. Eigenstates with outstanding values of $\beta$ are indicated by stars. 
}  
\label{trimersigma}  
\end{figure}

\begin{figure}

\includegraphics[width=6cm]{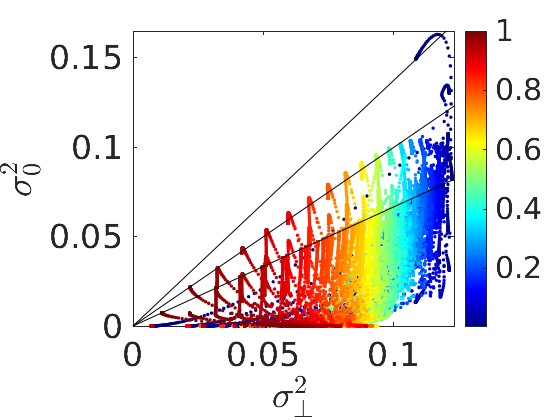}

\includegraphics[width=6cm]{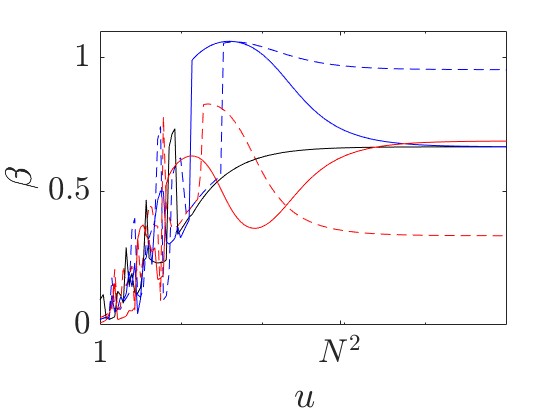}	

\includegraphics[width=6cm]{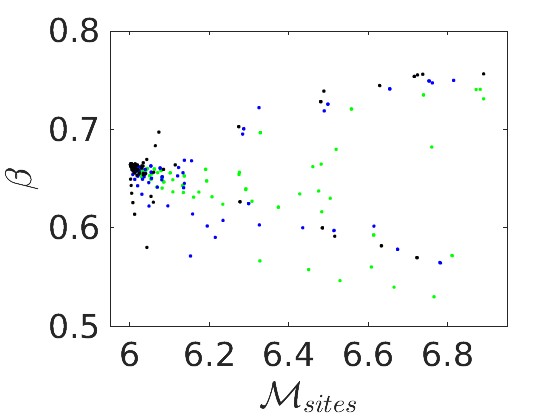}

\caption{
{\bf Identification of amplitude modes.} 
Upper panel: scatter diagram of $\sigma_\perp^2$ versus $\sigma_0^2$ for all the eigenstates in the diagrams of \Fig{trimersigma}. The color code is $E$. Points of lower-energy states are plotted last, and therefore may hide points of higher-energy states.  The straight black lines indicate the slopes ${\beta=2/3,1,11/8}$. 
Middle panel: Plot of $\beta$ versus $u$ for representative levels ${\nu=388,391,408,412,418}$. 
Roughly, they represent 5 families, where the dependence is either monotonic (black line) or non-monotonic. In the latter case $\beta$ is either larger or smaller compared with the generic MI prediction. 
Lower panels: $\beta$ vs $\mathcal{M}_{sites}$ for ${u=450,850,1700}$ (green, blue, black). This panel indicates that outstanding values of $\beta$ emerge as the MI levels start to mix.
}  
\label{HMtrimer} 

\end{figure}

The Higgs measure $\beta$ of \Eq{eHiggs} is the ratio of $\sigma_0^2$ to $\sigma_{\perp}^2$.  By definition it becomes of order unity for amplitude oscillations of the ODLRO.  Irrespective of that we have $\sigma_{\parallel}^2$ that characterizes the ``diagonal" on-site fluctuations.
Numerical results for the $\sigma$-s are summarized in \Fig{trimersigma}, and the result for the Higgs measure were already displayed in \Fig{trimerTomog}. We identify that there are levels where the dependence of $\beta$ on $u$ is non-monotonic. This is further illustrated, for representative levels, in \Fig{HMtrimer}. 
Below we clarify analytically the numerical results for the various families of eigenstates. 

\subsection{Generic MI states}

In this subsection we show that in the MI phase we get for all the {\em generic} eigenstates  $\beta$ of order unity. The term ``generic" requires clarification. Each MI eigenstate is a translation-invariant superposition of Fock states that have the same unperturbed energy. All the permutation of a given configuration have to appear in such superposition. What we call ``generic" MI state assumes that the pertinent configurations in the superposition differ from each other by more than 2 particle transitions. 
For such generic superpositions, say ${\Psi = \Psi^{(1)}+\Psi^{(2)}}$ we can use 
$\braket{A} = \braket{A}_1 + \braket{A}_2$   
for any one-body or two-body operator, 
because $\BraKet{\Psi^{(2)}}{A}{\Psi^{(1)}} = 0$.
Assuming a generic MI state that is characterized by occupations $n_j$ we get for the on-site fluctuations  
\beq
\sigma_{\parallel}^2 \ = \ \text{Var}(n_{site})  \ = \ \frac{1}{L} \sum_j n_j^2 - \left( \frac{N}{L} \right)^2
\eeq
The zero-momentum orbital occupation operator is 
\beq \label{enocc}
\hat{n}_0  =  b_0^{\dagger}b_0 
= \frac{1}{L}\sum_{i,j} a_i^{\dagger}a_j
= \frac{N}{L} +  \frac{1}{L}\sum_{i \neq j} a_i^{\dagger}a_j
\eeq
We get that $n_k= \braket{\hat{n}_k} =N/L$ for any $k$, while 
\beq \label{eSigG}
\sigma_{0}^2 \ &=& \ \text{Var}(n_0) 
\ = \ \frac{1}{L^2} \sum_{i \neq j} \braket{a_i^{\dagger}a_j a_j^{\dagger}a_i} 
\nonumber \\
\ &=& \ \frac{1}{L^2} \left[ 2N + N^2 - \sum_j n_j^2  \right]
\eeq
The ODLRO fluctuations can be extracted from the sum rule \Eq{eSR}, leading to  
\beq
\sigma_{\perp}^2 \ = \ \frac{1}{[2]L} \left[ (L{-}1)N + N^2 - \sum_j n_j^2 \right]
\eeq
where the factor $[2]$ should be omitted for the dimer. 

The Higgs measure $\beta$ is defined as the ratio between the amplitude fluctuation which are typically  
${\sigma_{0}^2 \sim (N/L)^2}$ and the total ODLRO fluctuations which are typically ${\sigma_{\perp}^2 \sim N^2}$.  Therefore $\beta$ comes out of order unity. Specifically we get ${\beta \approx 1/2}$ for the dimer, and ${\beta=2/3}$ for the trimer.

\subsection{Outstanding states in the MI phase}

The order parameter $n_0=\braket{\hat{n}_0}$ equals $N/L$ for any generic MI state. This is not the case if the eigenstate is a superposition that involves {\em different} Fock states that are coupled by an $a_{j}^{\dag}a_{i}$ ``transition terms" of the $\hat{n}_0$ operator, see \Eq{enocc}.  Such state belongs to an energy level that strictly speaking does not undergo an SF-to-MI transition. 

In the case of commensurate dimer all the levels undergo SF-MI transition, as opposed to the incommensurate dimer where the pair of ground-state levels does not undergo this transition: see \Fig{dimerHn}, where the ground state levels maintain ``polarization" also in the MI phase. 
But in the trimer, an inspection of the $n_0$~diagram (\Fig{trimerTomog}) shows that there are many levels that do not exhibit an SF-MI transition. These are levels that are formed of degenerate states that differ by a single-particle transition, i.e. permutations of ${\ket{n,\, n{+}1, \, N{-}2n{-}1}}$.   

There are also outstanding energy-levels that do undergo an SF-to-MI transition, but nevertheless feature a non-generic value of $\beta$. In the commensurate dimer case, only the two lowest excitations exhibit outstanding value of~$\beta$, see  \Fig{dimerE}. All the higher dimer excitations are generic, because the occupations of the two sites differ by more than two particle. As opposed to that, MI states with outstanding $\beta$ are rather frequent in the trimer spectrum, as can be seen from the $\beta$~diagram of \Fig{trimerTomog}, and from \Fig{HMtrimer}.

The value of $\beta$ for an MI state becomes non-generic if the superposition involves different Fock states that are coupled by $a_{i'}^{\dag}a_{j'}a_{j}^{\dag}a_{i}$ transition terms of the $\hat{n}_0^2$ operator (with ${j \ne i}$ and ${j' \ne i'}$). Such terms can enhance or suppress the dispersion $\sigma_0$. The detailed calculation of $\beta$ is explained in \App{aI}. Here we summarize the main outcomes.

The 6~lowest excitations of the commensurate trimer deserve special attention. They are superpositions of ${\ket{ \bar{n},\, \bar{n}{+}1, \, \bar{n}{-}1} }$ and its permutations. The lowest state in this sub-space is  a zero-momentum superposition with equal coefficients. It is polarized, with ${\braket{n_0}=(5/9)N}$ instead of ${\braket{n_0}=(1/3)N}$. Consequently $\sigma_{\perp}^2$ is somewhat reduced by factor $8/9$. But the main issue is the enhancement in $\sigma_0^2$, which is enhance by a factor $11/6$. Thus we deduce the non-generic exceptional ratio ${\beta \approx 11/8}$. This finding is very pronounced in the upper panels of \Fig{HMtrimer}.

Let us look in the other end of the spectrum. Consider the 6 trimer excitations that  are superpositions of ${\ket{N{-}1,\, 1, \, 0} }$ and its permutations. The lowest state in this sub-space is a zero-momentum superposition with equal coefficients. The order parameter $\braket{\hat{n}_0}$ is barely affected. In the $\sigma_0^2$ calculation, one observes that the enhancement factor is $6/4$. Consequently we deduce that this state features the non-generic value ${\beta \approx 1}$. This prediction turns out to be satisfactory for all the high-energy outstanding excitations, as observed in \Fig{HMtrimer}.

\subsection{Higgs at the MI-SF transition}

From the previous subsections we can deduce the variation of $\beta$ as we move from the SF regime to the MI regime. For the squeezed SF ground state, see \App{aB}, the amplitude fluctuations are ${\sigma_{0}^2 \approx 2\left(1+n_{dep}\right)n_{dep} }$, where $n_{dep}=N-n_0$ is the average depletion. 
The ODLRO fluctuation can be derived from the sum rule. Namely, the right hand side of \Eq{eSRt} implies \Eq{eSRSF}, if we assume that the depletion is small and ignore the residual minimum uncertainty. If we further neglect $\sigma_{\parallel}$, we deduce that for eigenstates in the vicinity of the SF ground-state ${\sigma_{\perp}^2 \sim Nn_{dep}}$, meaning that it is proportional to the depletion.  Consequently, we get for those states   
\beq
\beta \ \ \sim \ \ \frac{\text{Var}(n_{dep})}{Nn_{dep}}  
\ \ \sim \ \ \frac{n_{dep}}{N}
\eeq
%
After we cross to the MI phase, $\beta$ becomes of order unity. Specifically we get ${\beta \approx 1/2}$ for the dimer, and ${\beta=2/3}$ for generic MI eigenstates of trimer. For non-generic MI states we get either enhanced or suppressed value of $\beta$ as discussed in the previous subsection. 

Further inspection of \Fig{HMtrimer} reveals that there are levels that exhibit an outstanding value of $\beta$ only at the vicinity of the SF-MI transition. This is similar to what we have observed for the dimer in \Fig{dimerE}, but much more pronounced. The location of the conspicuous ${\beta\approx 1}$ eigenstates are indicated by stars in the third panel of \Fig{trimersigma}. It is natural to suggest that their appearance reflects mixing of level, as if the energy-band becomes effectively non-generic with larger quasi-degeneracy. This suggestion is confirmed by lowest panel of \Fig{HMtrimer}.

\section{Zooming into the lowest excitation band}
\label{s10}

\begin{figure}[t!]

\includegraphics[width=4cm]{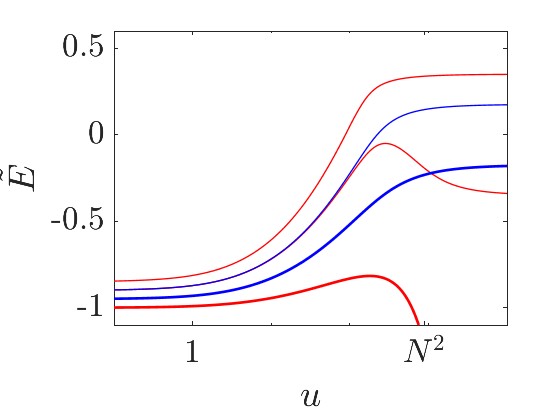}
\includegraphics[width=4.5cm]{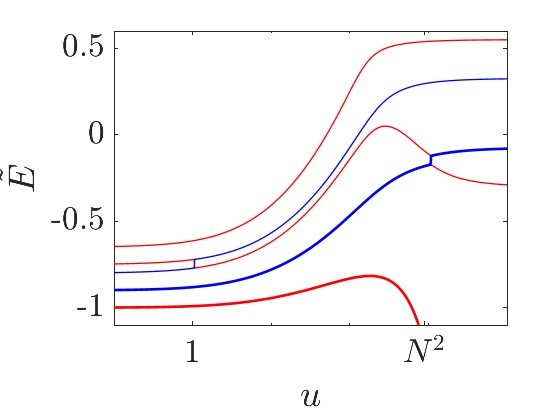} 
\\
\includegraphics[width=4cm]{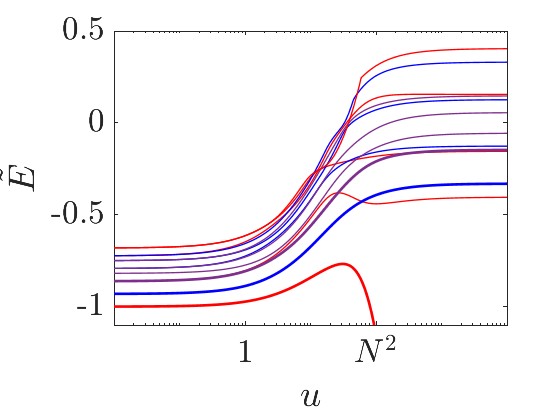} 
\includegraphics[width=4.5cm]{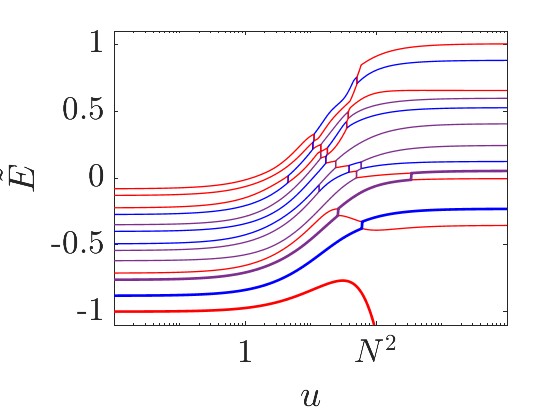} 

\caption{
{\bf The lowest energy levels.} 
The plots show how the ground-state and the first-band of MI states evolve as $J$ is increased (smaller $u$), highlighting the formation of a Goldstone and Higgs bands. Vertical axis is $\tilde{E}=(E-E_b)/(NJ)$, where $E_b$ is the unperturbed ($J=0$) energy of the first band. Upper panels are for the trimer with $N=30$.
Lower panels are for $L{=}5$ ring with $N=10$.
Red lines indicate ${q=0}$ levels. 
The right panels are obtained from the left panels 
by adding vertical spaces between levels,   
as to achieve better resolution of their order. 
The ground state and its Goldstone excitations are indicated by thicker lines (blue and violate lines that migrate from the upper MI band).
}  
\label{lowLevels}  
%
%
%

\ \\

\includegraphics[height=4cm]{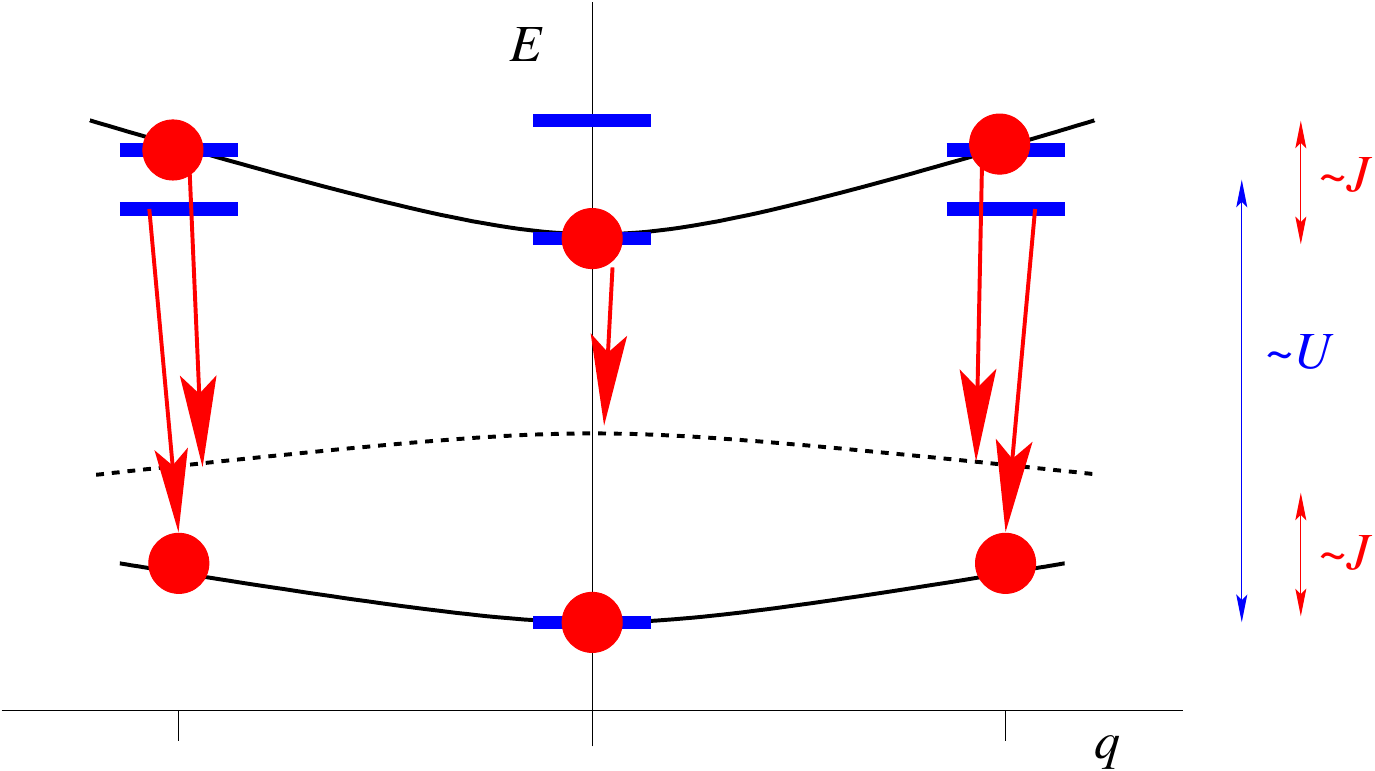} 

\caption{
{\bf The formation of bands.} 
This caricature clarifies how the MI bands of the trimer are re-arrange into SF bands, as $J$ is increased. In the MI phase we have two sub-bands of width $\sim J$, that form the first band of quasi-degenerate particle-hole excitations. This band has a gap $\sim U$ from the ground state. The associated eigen-energies are indicated by blue segments. As $J$ is increased, Goldstone and Higgs bands are formed. Their eigen-energies are indicated by red circles. For much larger $J$ the Higgs band migrates into the SF region, and becomes a non-gapped second band of two-phonon excitations (dotted line).           
}  
\label{lowLevelsSc}  
\end{figure}

\Fig{lowLevels} displays the lowest energies $E_{\nu}$ versus $u$ for the $L=3$ trimer and for an $L=5$ ring. The $q=0$ levels are colored in red, and in some sense ``frame" the band structure. There are two limits where the structure of the spectrum is rather simple. 
In the {\em MI phase} we have the $q{=}0$ ground-state, and the first band of excitations that contains $L{-}1$ sub-bands, each with $L$ momentum states ${q=(2\pi/L)\times \text{integer}}$. Those states are superpositions of gapped particle-hole excitations. 
In the {\em SF phase} we have the $k{=}0$ ground state, the single-phonon band that contains $L$ states, and the double-phonon states that contain both $q=0$ states that are formed from $\pm k$ excitations, and $q\ne0$ double-phonon excitations. 

The non-trivial aspect is the {\em MI-SF transition}, during which there is migration of $q{\ne}0$ levels from the gapped MI band towards the $q{=}0$ ground level, leading to the formation of a Goldstone band in the SF region. This migration is caricatured in \Fig{lowLevelsSc}.

We now turn to provide a more detailed semiclassical description of the MI-SF transition. The lowest seperatrix defines the SF region ${E<E_{SF}}$. The energy width of this region is of order $NJ$. The quantum parameter $\gamma$ of \Eq{eGamma} tells us what is the  size of Planck cell relative to the size of the SF region. Large value (${\gamma > 1}$) implies that the SF region cannot accommodate quantum eigenstates. This is the MI phase. This phase features narrow bands of particle-hole excitations, see \Fig{lowLevelsSc}. The spacing between those bands is of order~$U$.  Due to $J$ they form sub-bands. As $\gamma$ becomes smaller eigenstates migrate  through the separatrix into the SF region, where they are re-arranged into phononic bands. The latter can be regraded as occupation states of momentum-orbitals.  

The levels that migrate into the SF region form the so-called Goldstone band. The latter has no gap from the ground state (it is like small vibrations around the ground state). As opposed to that, the MI gapped bands are formed of levels that reside above the separatrix. For large enough $J$, the SF regions expands and swallows all the eigenstates, and accordingly all the bands become phonon-type.

Let us take a closer look at a trimer that is occupied by $\bar{n}$ states in each site. The first band of excitations is spanned by the 6 basis states $|x,s\rangle$ of \Eq{eFockS}, where ${x=1,2,3}$ is the ``position" index of the configuration, and ${s=a,b}$ is like a sublattice index that distinguish two sets of configurations that differ by the cyclic order of particle-hole excitations. Upon translation ${D |x,\sigma \rangle = |x{+}1,\sigma \rangle}$~mod(3). The hopping terms are alternately $-(J/2)[1{+}\bar{n}]$ and $-(J/2)\bar{n}$. The eigenstates of the Hamiltonian form two bands. By Bloch theorem
\beq
|q,\pm\rangle = \sum_x  e^{iqx} \left[ a_{\pm} |x,a\rangle + b_{\pm} |x,b\rangle  \right]  
\eeq
with quasi-momentum ${q=2\pi/3 \times \text{integer}}$, namely, ${  D|q,\pm\rangle = e^{-iq} |q,\pm\rangle }$. The lowest excitation ${|q{=}0, +\rangle}$ is a zero momentum particle-hole excitation that features the outstanding large Higgs measure ${\beta \approx 11/8}$, as derived in \App{aI}. Note that the excitations in this band are insensitive to the introduction of external gauge field.

\section{Summary and discussion} 
\label{s11}

We have presented a semiclassical tomographic approach to study the
superfluid-insulator transition in finite-size Bose-Hubbard rings.
Various measures have been introduced to inspect the many-body eigenstates, 
and signatures of the underlying mixed chaotic dynamics have been identified.  
A special attention has been devoted for the characterization of fluctuations. 
Below we expand on several perspectives that have motivated this work.

\subsection{Atomtronic perspective}

The formation of one-dimensional lattice rings is state of the art, see \cite{atomtronics} and references therein. The bosons can be trapped by a painted optical potential as demonstrate (e.g. in Fig.11 there). 
The strength of the interaction ($u$) can be tuned, and remaining challenge is to define a protocol for the preparation of energy $E$ states. Optionally the ring can be rotated (gaining control over $\Phi$).
The occupation of the momentum orbitals and the current can be measured from time-of-flight images (see e.g. Fig.15 there). It is natural to adopt tomographic measures that are related to such measurements.  See for example discussion of density-density covariance measurements in Section~X-C there. To be precise, we are dealing here with counting-statistics.

\subsection{Semiclassical perspective}

The emergence of chaos for rings of different size ($L$) with different particle number ($N$) has been studied in \cite{BHHchainChaos2}. The $(u,E)$ phase-diagrams have been found numerically for rather large rings based on spectral statistics (the $r$ measure) and fractal dimension (essentially $\mathcal{M}$) of the eigenstates. Disregarding quantitative aspects it is fair to say that those numerically-demanding diagrams are rather similar to those of the $L{=}3$ trimer that we have analyzed in the present work. 

However, our purpose was not just to look for ``chaos", but to provide a detailed inspection of the SF-MI transition using a refined set of measures that allow ``tomography" of the underlying phase-space. Let us discuss whether analysis of ${L{>}3}$ rings would be merely ``more of the same" or whether it is likely to provide ``new" physics that is absent in the ${L{=}3}$ model. Before we further discuss this question, we have to emphasize the following: (a)~Our approach is both numerically efficient and informative; (b)~In principle our approach can be used for larger rings. The second point is non-trivial. The use of Poincare sections is feasible only for systems with two degrees of freedom. True that we have generated such sections to support the phase-space interpretation of some findings, but this should not be regarded as part of the ``tomography" package. As far as tomography is concerned, each eigenstate is characterized by a set of {\em measures} whose definition does not assume that we are dealing with an ${L{=}3}$ ring. Therefore, we can carry out the same type of tomography for any ${L>3}$ ring.                             
 
Systems that have $d{=}2$ degrees of freedom, such as the ${L{=}3}$ trimer, are in some sense special. We can say that they feature low dimensional chaos: they have phase-space with distinct quasi-regular and chaotic regions. This is because the energy surface has ${2d{-}1=3}$ dimensions, while KAM barriers has $d{=}2$ dimensions. If we have $d{>}2$ system the KAM barriers cannot divided the energy surface into distinct regions. Even if the non-linearity is small there is always slow leakage, aka Arnold diffusion, via dynamical barriers. It follows that in some sense we always have full chaos (as opposed to mixed phase-space). Consequently one may speculate that ${L=3}$ rings are distinct compared to ${L{>}3}$ rings. But in a quantum context it turns out that the identification of this distinction requires very special type of experiments, as discussed in \cite{mbp}. As far as eigenstate tomography is concerned the distinction between low and high dimensional chaos is blurred  for complementary reasons: (a)~In low dimensional phase space quantum dynamics allows leakage via dynamical barriers; (b)~In high dimensional phase space quantum dynamics does not allow Arnold diffusion via small openings; (c)~Most quantum eigenstates display hybrid features as discussed in \cite{sst}. Consequently we do not expect any dramatic differences between the phase-diagrams of small and large rings. Difference are expected to be merely quantitative.   

The major borders in the ${(u,E)}$ phase-diagram of the BHM are separatrices ${(E_{SB}, E_{SF}, E_{ST})}$ and the perturbative border $u_s(E)$. The ODLRO and the associated fluctuations are ``imaged" on this diagram. In particular it was important for us to clarify the notion of MI-SF transition in the context of such diagram, and not to focus only on the ground-state. It is important to realize that the transition has several stages. As $u$ is decreased, we first approach the perturbative border $u_s(E)$ where fluctuations become outstanding. After crossing this border the eigenstates migrate to the SF region below the seperatrix. Whether a mobility edge is formed in the thermodynamic limit remains a matter for speculations.

\subsection{GMFT perspective}

The study was partially motivated by the desire to make a bridge with the field-theory perspective, notably with GMFT \cite{GMFTr1,GMFT1,GMFT2} that has been used to study time dependent scenarios, possibly with disordered lattices as in \cite{GMFTr2,GMFTr3}. In particular such framework has been used \cite{GMFT3} for the identification of amplitude (Higgs) modes as opposed to phase modes. 
Let us recall  how phase and amplitude modes are defined within this common perspective. It is natural to start with the familiar Bogulyubov framework, where the classical variation of the order parameter is described by  
\beq
\psi_j(t) &=&  \sqrt{n_j(t)} e^{i\varphi_j(t)} 
\nonumber \\
&=& \psi^{(0)} + u  e^{i(kx_j{-}\omega t)}  - v^* e^{-i(kx_j {-} \omega t)}  
\eeq
The variation of $\psi$ in the complex plane is along an ellipse that is characterized by a squeeze factor ${ v/u=\tanh(\lambda) }$. Using the common gauge convention, $\psi^{(0)}$ is real and positive, and both $v$ and $u$ are real numbers. In the standard Bogulyubov framework both have the same sign, and ${|v/u|<1}$. 

The GMFT suggests {\em quantum} results for $v/u$, that go beyond the Bogulyubov prediction. One considers the dynamics of a single site under the influence of a frozen mean-field of the other sites. Such approximation automatically excludes irregular chaotic motion. Nevertheless, for the regular solution it predicts the appearance of pure amplitude (Higgs) modes as the $J/U$ ratio is increased. 
{\em The argument goes as follows:} 
For $J=0$ the excitations are either {\em particle} or {\em hole} depending on the chemical potential. 
In the latter case it formally means $v/u=\pm\infty$. 
Setting ${J \ne 0}$, and considering a hole excitation, the mode changes its character as $J$ is increased, and for large enough $J$ it eventually becomes a particle-like Bogulyubov excitation with ${0 < v/u < 1 }$. It follows that along the way we should encounter either pure phase-mode (formally $v/u=+1$)  or pure amplitude mode  (formally $v/u=-1$). The latter is termed Higgs mode, and constitutes an  indication of a quantum interference effect that goes beyond the Bogulyubov framework.

The present work, unlike the GMFT, treats the full many-body problem. We note that a parallel study \cite{WG} aims to develop a theory for the Goldstone and Higgs excitations based on a refined cumulant expansion around the self-consistent mean field. Nevertheless, in the exact numerical treatment, it is $N$ rather than the chemical potential that is fixed. Accordingly the lowest excitations in the MI phase are correlated particle-hole excitations. This complicates the practical comparison between field-theory predictions and  numerical results.    
%
%
Consequently, we have adopted a theoretically unbiased characterization of eigenstates that reflects the distinction between phase and amplitude modes. Specifically, we have defined an Higgs measure $\beta$ to identify eigenstates that have outstanding amplitude fluctuations.  

For the dimer, the observed picture is very simple: eigenstates that feature outstanding $\beta$ appear at the MI-SF transition when an even-symmetry eigenstate crosses the separatrix. It is complementary to the localization of the eigenstate at the unstable hyperbolic fixed-point.    
For the trimer the picture is more complicated because the MI-SF transition is mediated by chaos. Nevertheless, we found that outstanding $\beta$ is not related to separatrix crossing, but rather can be explained within the framework of perturbation theory. Namely, it is related to special superpositions of quasi-degenerate eigenstates.

\ \\ \ \\ 
{\bf Acknowledgments} --  
We thank Idan Wallerstein and Eytan Grosfeld for insightful discussions. Preliminary work that concerns \Sec{s10} has been carried out by Naama Harcavi within the framework of a BSc project. The research has been supported by the Israel Science Foundation, grant No.518/22.

\appendix

\newpage
\section{The generalized Bloch vector}
\label{aA}

We define ${S_{i,j} \equiv a_i^{\dag} a_j}$. 
For ${i \neq j}$ we use the optional notation 
$S_{i,j} \equiv \hat{S}_{x}+i\hat{S}_{y}$.   
The subscript $x$ will be used below as a bond index. In an $L$ site system we have $(L{-}1)L/2$ bonds. 
The $S_x$ and $S_y$ operators can be regarded as the components of the ``order parameter". If they have a non-zero expectation value, it is implies that the system is ordered (phase-correlated)
The expectation values $\braket{S_{i,j}}$ are the components of a generalized Bloch vector. The diagonal elements of $\braket{S_{i,j}}$ are the site occupations.  The off-diagonal elements of $\braket{S_{i,j}}$ provide an indication for ODLRO. 
Assuming clean ring with translation symmetry,  
we always have ${\braket{S_{i,i}} =N/L}$. 
The ODLRO is related to the occupations 
$n_k \equiv \braket{\hat{n}_k} $
of the momentum orbitals.
We have the relation  
\beq \label{eSr}
S(r) \ = \ \frac{1}{L} \sum_k n_k \, e^{-ikr} 
\ \equiv \ S_x(r) + i S_y(r)
\eeq
Here ${r=(i-j)}$ is a dummy index that reflects the ``distance" between the sites. Full condensation in the zero momentum orbital implies full ODLRO with ${S(r) = N/L}$. 
More generally, we get the following relations 
\beq
S(1)_{dimer} &=& \left(n_0{-}\frac{N}{2}\right) 
\\
S(1)_{trimer} &=& \frac{1}{2}\left(n_0{-}\frac{N}{3}\right) -i \frac{\sqrt{3}}{6} (n_{+}{-}n_{-}) 
\\
S(r {\neq} 0)_{general} &\sim& 
\frac{1}{L} [N - \alpha (N{-}n_0)]  
\eeq
where $\alpha$ is numerical prefactor.

Above we have expressed the $S_{i,j}$ in terms of the $n_k$. An inverse relation relates the occupation of the zero-momentum orbital to the components of the order parameter: 
\beq 
\hat{n}_0 
\ \ = \ \ \frac{N}{L} + \frac{2}{L}\sum_{x}^{(L{-}1)L/2} S_x
\eeq
Therefore 
\beq 
n_0 \ \ &=& \ \ \frac{N}{L}  + \sum_{r=1}^{(L{-}1)} S_x(r) 
\eeq
It is important to realize that the fluctuations of the zero momentum orbital reflects only the fluctuations of the $S_x$ components of the order parameter. Accordingly $\sigma_0^2$ reflects {\em amplitude} fluctuations of the order parameter, as opposed to phase fluctuations that are reflected by the variance of $S_y$.

\section{Purity and depletion}
\label{aB}

For the dimer it is common to use spin-language. For each eigenstate we calculate the Bloch vector ${\vec{S}=(\braket{S_x},\braket{S_y},\braket{S_z})}$, which is merely a representation of the reduced probability matrix. This determines the location ${ (\theta,  \varphi) }$ of the eigenstate in spherical coordinates on the Bloch sphere.
In the presentation of the main text, the dimer ground-state is a squeezed state that is oriented in the ${(\theta=\pi/2,\varphi=0)}$ direction. The squeezing is in the $\theta$ direction, while the stretching is in the azimuthal direction $\varphi$.
In this appendix, without loss of generality, we assume that the axes of the Bloch sphere are rotated such that ${ \theta=0 }$. With this convention, the operator that destroys excitations is 
\beq
A \ \ = \ \  \frac{1}{\sqrt{N}} S_{+}   \ \ \equiv \ \ \frac{1}{\sqrt{2}}(q+ip)  
\eeq
Locally at the vicinity of ${ n\sim 0 }$ we have algebra of harmonic oscillator. We define a depletion coordinate
\beq
\hat{n}  = \hat{n}_{dep} = \frac{N}{2}-S_z  \approx  A^{\dagger}A 
\eeq
The latter approximation assumes ${ n \ll N }$. For the purity we get the relation  
\beq
\mathcal{S}  \ \ =  \ \   \frac{1}{2}\left[1 + \left(\frac{2}{N}\right)^2 |\vec{S}|^2\right]   
\ \  \approx \ \  1 - \frac{2}{N} \langle \hat{n} \rangle  
\eeq
We can define a covariance matrix in the ${ (q, p) }$ coordinates. 
By an appropriate rotation we can get zero for the correlation 
and $\sigma_q^2$ and  $\sigma_p^2$ for the variances.  
We realize that
\beq
\langle \hat{n} \rangle \ \ \approx \ \ \frac{1}{2} (\sigma_q^2 + \sigma_p^2 -1)  \ \ \equiv \ \  [\sinh(2\lambda)]^2
\eeq
where the latter equality defines the squeezing factor. 
Following \cite{Squeezed} we have 
\beq
\sigma_0^2 \ &=& \ \text{Var}(\hat{n}) \ = \ 2[\sinh(2\lambda)]^2[\cosh(2\lambda)]^2 
\nonumber \\
\ \ &=& \ \  2\left(1+\braket{\hat{n}}\right)\braket{\hat{n}}
\eeq
For completeness we note that the squeeze factor $\lambda$ of the low excitations can be calcualated as follows
\beq
\lambda \ \ \approx \ \ -\frac{1}{2} \ln \sqrt{\frac{\braket{S_x^2}}{\braket{S_y^2}}} 
\eeq
while the squeeze operation is 
\beq
[S(\lambda)]^{-1} &=& e^{-\lambda(A^2-(A^\dagger)^2)/2} = e^{-\lambda(S_{+}^2-S_{-}^2)/(2N)} 
\nonumber \\
&=& e^{-i\lambda(S_xS_y+S_yS_x)/N}
\eeq

The above approximated relation between the purity and the depletion can be generalized to multi-site systems.  We first select an orbital basis such that the reduced one-body probability matrix ${ \rho }$ is diagonal. The orbitals are indexed ${ k = 0, 1, 2, \cdots }$.  
We define field operators indexed by ${ k=1,2,\cdots }$,
\beq
A_k  \ \ = \ \  \frac{1}{\sqrt{N}} a_0^{\dagger} a_k   \ \ \equiv \ \ \frac{1}{\sqrt{2}}(q_k+ip_k)  
\eeq
From the expectation values of $A_kA_k^{\dagger}$ we can calculate the $n_k$ and get ${ n_0 = N {-} \sum_k  n_k }$. 
Then, from ${ \rho = (1/N) \text{diag}\{  n_0,  n_1, \cdots \}  }$, one obtains 
\beq
\mathcal{S}    \ \  \approx \ \  1 - \frac{2}{N}  \sum_{k\ne 0}  n_k  \ \ = \ \ \frac{2}{N} n_0 - 1  
\eeq

\section{Husimi functions}
\label{aC}

We use the symbol $\alpha$ to indicate a point is phase-space of a single particle system. For harmonic oscillator ${\alpha = (q,p)}$ represents a pair of conjugate coordinates, while for a dimer we use Bloch sphere coordinates  ${\alpha = (\theta,\varphi)}$. For a BHM with more than 2 sites the generalization is straightforward, 
and requires two $\theta$-s to indicate relative site occupations, and two conjugate $\varphi$-s to indicate relative phases.  
In a quantum context $\alpha$ actually specifies an orbital that can accommodate particles. The associated creation operators are: 
\beq
c_{\alpha}^{\dagger} \ = \ \sum_j C_j e^{i \varphi_j} a_j^{\dagger}, 
\ \ \ \ \ \text{with} \ \ \sum |C_j|^2=1
\eeq
Consequently, a manybody coherent-state is defined as follows
\beq
\ket{\alpha} \ \ = \ \ \frac{1}{\sqrt{N!}}\left[c_{\alpha}^{\dagger} \right]^N \, \ket{\text{vacuum}}   
\eeq
For a dimer, the the standard basis is the site-Fock basis ${\ket{n}}$, where ${n=(N/2)-S_z}$, and the explicit representation of the coherent states is 
\beq
|\alpha \rangle = \sum_{n=0}^N \sqrt{\left(\amatrix{N \\ n}\right)} 
\left[\cos{\frac{\theta}{2}}\right]^{N{-}n} 
\!\left[\sin{\frac{\theta}{2}}\right]^{n}   
e^{i n \varphi} \ |n \rangle 
\ \ \ 
\eeq

The Husimi function use this over-complete basis of $\alpha$ states in order to represent the many-body quantum state. Namely, it is defined as follows:  
\beq \label{eHus} 
Q(\alpha) = |\langle \alpha | \Psi \rangle |^2
\eeq
For a dimer the function is plotted on the two dimensional Bloch sphere.  For a trimer we have to select a section. This can be done in one-to-one correspondence with Poincare section. It is customary to plot the the section at energy $E$ that equals the eigenstate energy $E_{\nu}$. In such procedure forbidden regions are {\em excluded by construction}.  

We use an optional procedure for plotting a modified quantum Poincare section.  We write the `position' coordinate as ${ \bm{n}=(n_1,n_2) }$. It is implicit that ${n_3=N-n_1-n_2}$. We define a reduced wavefunction 
\beq
\psi_n \ \ = \ \  \sum_{ \bm{n} \in n } \langle \bm{n} | \Psi \rangle
\eeq
%
The summation is over the excluded coordinate ${ n_2 }$, keeping the distinguished coordinate  ${ n_1 = n }$ fixed. It is like selecting the wavefunction amplitudes for which the excluded coordinate has zero momentum. Then we normalize $\psi_n$ and calculate the Husimi function as if we were dealing with a dimer. We call the outcome {\em Quantum Poincare Section}, as to distinguish it from the common {\em Husimi function}. The possible disadvantage of the Quantum Poincare Section is as follows: it corresponds to the union of two branches of the classical Poincare section.  Namely, in the classical definition one keeps points of the trajectory that cross the section in one selected direction.

\begin{figure*}
\includegraphics[width=16cm]{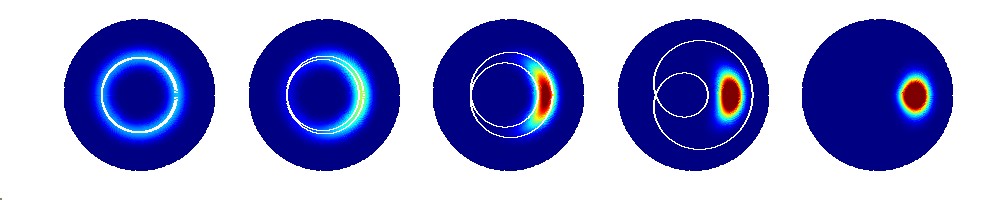}
\includegraphics[width=16cm]{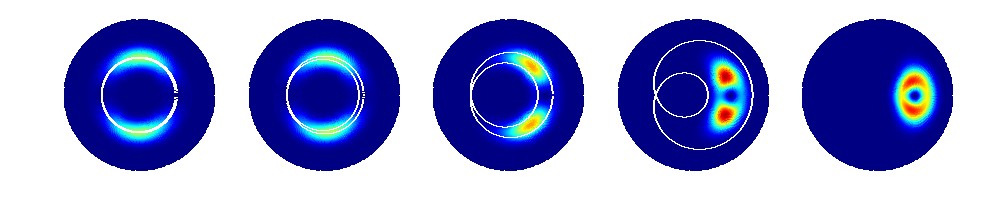}
\includegraphics[width=16cm]{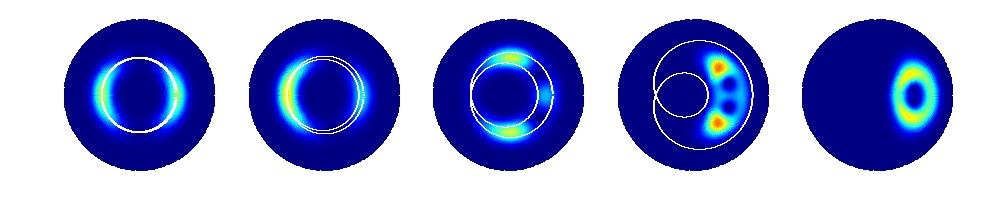}
\includegraphics[width=16cm]{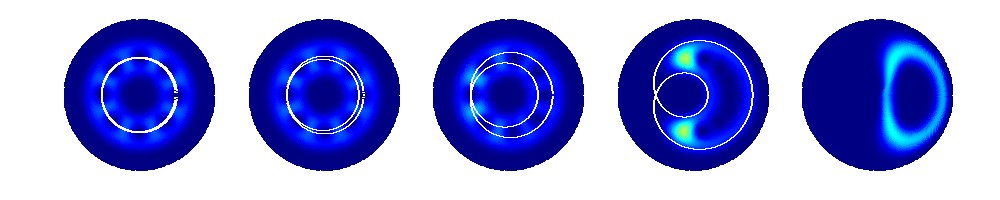}
\caption{
{\bf The MI-SF transition for the dimer eigenstates.}
The number of bosons is $N=30$.
The rows are for the levels $\nu = 0,1,2,7$. 
The columns, right to left, 
are for $u=1,10,100,10^3,10^4$. 
The coordinates in each panel are $(n_{site},\varphi_{site})$,
such that the Origin is the North pole, 
while the perimeter is the South pole.
}  
\label{dimerH}  
%
%
\includegraphics[width=16cm]{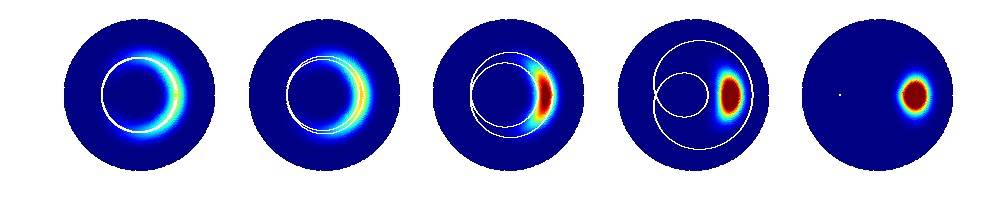}
\includegraphics[width=16cm]{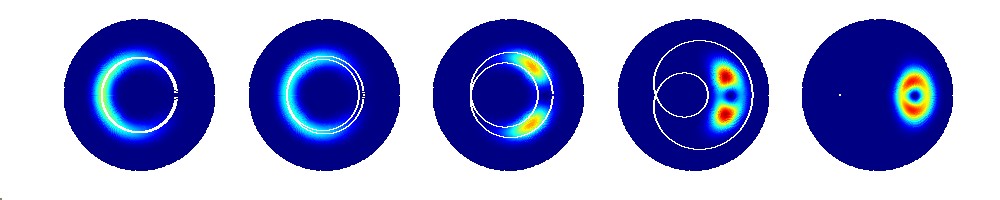}
\caption{
{\bf The MI-SF transition for an odd number of particles.}
The number of bosons is $N=31$.
The rows are for the levels $\nu = 0,1$. 
The columns, right to left, are for $u=1,10,100,10^3,10^4$. 
The coordinates are as in \Fig{dimerH}.
Here the low lying states are quasi-degenerate 
and analogous to the $\pm X$ polarization states of spin 1/2 entity. Both are even in the azimuthal coordinate. 
There remain polarized for arbitrary large $u$, meaning that there is no SF-to-MI transition. 
}  
\label{dimerHn}  
\end{figure*}

\section{Husimi functions for the dimer}
\label{aD}

In \Fig{dimerH} we display Husimi functions of the low lying eigenstates. We use stereographic projection such that the North pole is in the center, while the South pole is the outer circle. Optionally, we can say that we are using polar plot, such that the radial coordinate is essentially the occupation $n_{site}$ of one of the sites.  
The left most column is for large $u$. From left to right $J$ is increased, and consequently the MI ground state (GS) and its excitations evolve parametrically into the SF region in phasespace.

The migration of eigenstates from the ST-region to the SF region looks different for odd and even states. The GS has even symmetry. The MI excitations are the odd and even superpositions of unbalanced occupation states. The odd superposition evolves smoothly into the SF region, and settles around the stable fixed point. After that, the even superposition first {\em localize} at the hyperbolic point, and only after that migrates into the opposite SF region, settling around the stable fixed point. 

For completeness we show in \Fig{dimerHn} the parametric evolution of the ground state levels for an incommensurate dimer.  Those levels remain polarized for arbitrary large u, meaning that there is no SF-to-MI transition.

\section{The perturbative border}
\label{aE}

The unperturbed MI eigenstates, disregarding degeneracies,  are Fock site-occupation states $\bm{n}$, indexed by $\nu$, with energies 
${ E_{\nu}^{(0)}=(U/2)\sum n_j^2 }$.  These states are by the hopping. Assume that a particle hop from site \#1 to site \#2, the energy difference is 
${\Delta = U[1+(n_2{-}n_1)]}$, and the coupling is 
${W \approx J \sqrt{n_1 n_2} }$. We would like to obtain an estimate for $W/\Delta$. For the dimer it is straightforward, and the result can be written as 
\beq \label{eWoD}
\left|\frac{W}{\Delta} \right| \ \ \approx \ \ \frac{J}{U}\sqrt{\frac{E_{max}-E}{E-E_{min}}} 
\eeq
The condition  ${|W/\Delta| \sim 1}$ determines the perturbative border \Eq{eUs}. One observes that in the central energy range both $E_{max}{-}E$ and $E{-}E_{min}$ are of order $N^2U$ and therefore  ${|W/\Delta| \sim N/u}$. As opposed to that, close to the ground state $E{-}E_{min}$ becomes of order $U$ and therefore ${|W/\Delta| \sim N^2/u}$. From this one concludes that the breakdown of perturbation theory is at ${u \sim N^2}$ for the ground state, which signifies the MI-SF transition. But for the majority of states, in the central part of the spectrum, we get the breakdown at ${u \sim N}$.  

The above estimate \Eq{eWoD} is valid also for an $L$ site ring, in a statistical sense. This is based on the following identity:
\beq \nonumber 
&&\overline{n_i n_j} \ \ = \ \ \frac{1}{(L{-}1)L} \sum_{i\ne j} n_i n_j 
\\ \nonumber 
&& \ = \frac{1}{(L{-}1)L} \left[N^2 - \sum_j n_j^2\right] 
\ \sim \ \frac{1}{U} [E_{max}{-}E]
\eeq
The above can be used in order to estimate~$W$. Similar reasoning applies to the estimate of $\Delta$. Namely, exploiting the linear relation between $\overline{|n_i{-}n_j|^2}$ and $\overline{n_i n_j}$, the estimate for $\Delta$ is linear in $E$, and should become vanishingly small at the ground state, hence it is proportional to  $[E{-}E_{min}]$.

\section{The accessible space}
\label{aF}

\begin{figure}[t!]
	
\includegraphics[width=4cm]{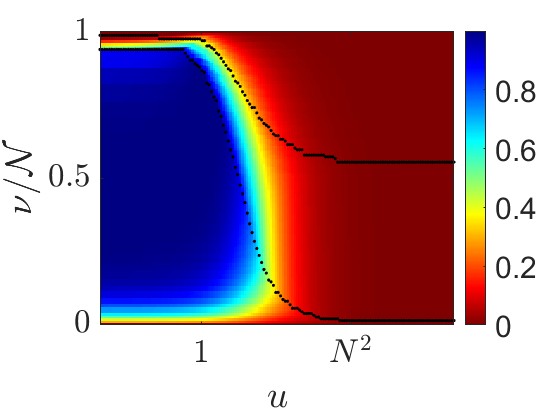}
\includegraphics[width=4cm]{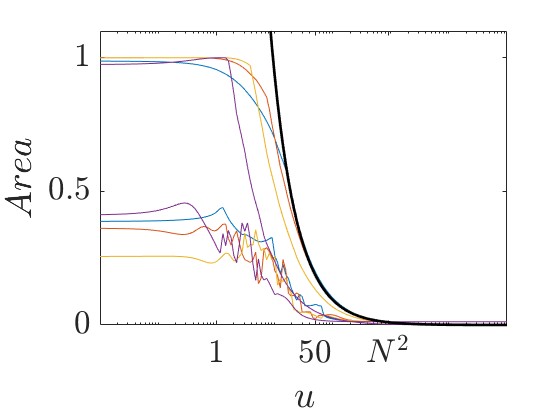}
	
\caption{
{\bf The accessible $\bm{n}$-region.}
The normalized accessible area $\mathcal{M}_s/\mathcal{N}$ is calculated as a function energy, showing the dependence on $u$.   
The left diagram is an image of the result, while the black line in the right panel shows the approximation ${3\pi/u}$. This is compared with $\overline{\mathcal{M}}_{sites}/\mathcal{N}$ (upper curves) and contrasted with $\mathcal{M}_{sites}/\mathcal{N}$ (lower curves) for the levels 
${\nu/\mathcal{N}=0.2,0.4,0.6,0.8}$.
}
\label{trimerArea}  
%
\ \\
%
\includegraphics[width=4cm]{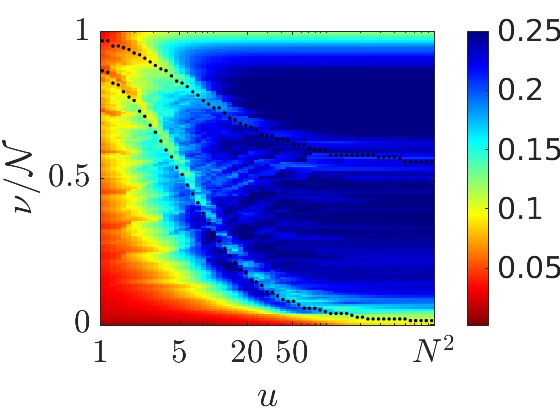}
\includegraphics[width=4cm]{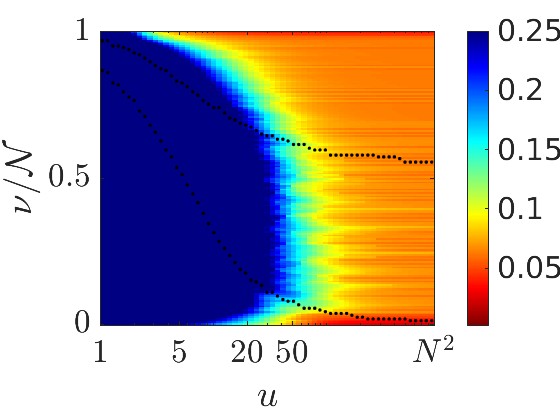}

\caption{
{\bf The participating $\bm{n}$-states.} 
These are extra panels for \Fig{trimerTomog}. 
The left and right panels display separately  $\overline{\mathcal{M}}_{orbital}/\mathcal{N}$
and $\overline{\mathcal{M}}_{sites}/\mathcal{N}$.	
}
\label{trimerMs}  
\end{figure}

The total area of $\bm{n}$-space is 
${\mathcal{N} \approx (1/2)N^2}$. 
The energetically accessible area $\mathcal{M}_s$ for the wavefunction $\Psi_{n_1,n_2}$  is the site representation is determined by the condition ${V_{-}(\bm{n}) <  E < V_{+}(\bm{n})}$. It is either a disc for ${E<E_{SF}}$, or an annulus for ${E_{SF}<E<E_{ST}}$, or fragmented for ${E>E_{ST}}$. The results of the numerical calculation of the accessible area as a function of ${(u,E)}$ are presented in the left panel of \Fig{trimerArea}. 

In the region where the accessible area is an annulus, it is easy to obtain an explicit expression. The energy width of the annulus is $(3/2)NJ$, and ${V(\bm{n})=(U/2)\sum n_j^2 }$ is a quadratic expression. Accordingly the accessible area comes out independent of $E$, namely, 
\beq
\mathcal{M}_s \ = \ \left(\frac{\pi}{U}\right) \frac{3}{2}NJ
\ = \ \frac{3\pi}{u}  \mathcal{N}
\eeq
In the right panels of \Fig{trimerArea} we compare this prediction with 
$\overline{\mathcal{M}}_{sites}/\mathcal{N}$. The agreement is satisfactory. Clealry, without the microcanonical averaging we get  $\mathcal{M}_{sites}$ that is much smaller than $\overline{\mathcal{M}}_{sites}$.

In the right panel of \Fig{trimerMs} we show the results for $\overline{\mathcal{M}}_{sites}$ in the full ${(u,E)}$ diagram, and clearly it resembles the left panels of \Fig{trimerArea}. In the left panel of \Fig{trimerMs} we show the results for $\overline{\mathcal{M}}_{orbitals}$. Both panels are merged in \Fig{trimerTomog}c. 

\section{The semiclassical border}
\label{aG}

In the vicinity of the ground state the trimer Hamiltonian can be written as ${\mathcal{H} = \mathcal{H}^{(0)} + \mathcal{H}^{(\pm}}$, see Eq(1) of \cite{qtp}, 
where ${ \mathcal{H}^{(0)} }$ is an integrable Hamiltonian whose constant of motion is the occupation imbalance ${(n_{+}{-}n_{-})}$. This constant of motion reflects that the main effect of the interaction is to create or destroy $\pm k$ pairs of particles in the excited orbitals. We call it {\em Bogoliubov approximation}. We emphasize that we avoid the further simplification of it into a quadratic form, which is the common practice in textbooks. The extra terms ${ \mathcal{H}^{(\pm)} }$ spoils the integrability and generate chaos. 

\begin{figure}

\includegraphics[width=4cm]{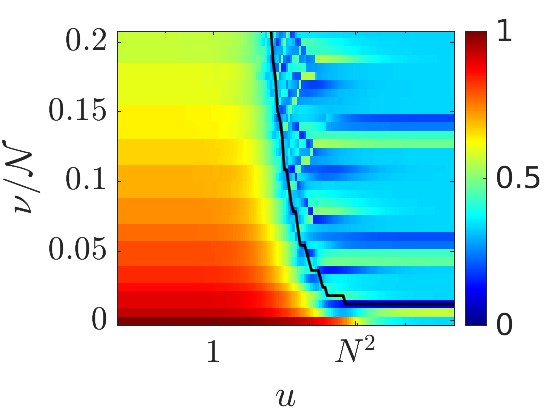}
\includegraphics[width=4cm]{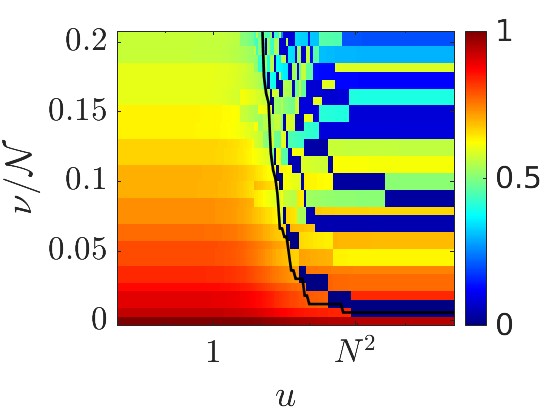}

\caption{
{\bf The trimer spectrum via Bogoliubov.}
The left panel is zoom over the lower part of the spectrum of \Fig{trimerSpect}, displaying~$n_0$. 
The right panels is the outcome of diagonalization of the Bogoliubov-approximated integrable hamiltonian. 
The line is $E_{SF}$. 
}
\label{trimerBogo}  
\end{figure}

In \Fig{trimerBogo} we show how the ${(u,E)}$ phase-diagram would look like if we ignored the chaos. It is obtained from the diagonalization of $\mathcal{H}^{(0)}$ (right panel) and compared with the diagonalization of $\mathcal{H}$ (left panel). We see that due to chaos the SF border is pushed to the left away from $E_{SF}$. This reflects the formation of an underlying stochastic region in the vicinity of the separatrix.   

\begin{figure}

\includegraphics[width=4cm]{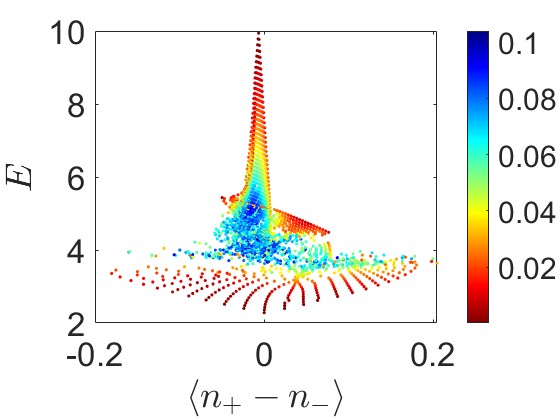}
\includegraphics[width=4cm]{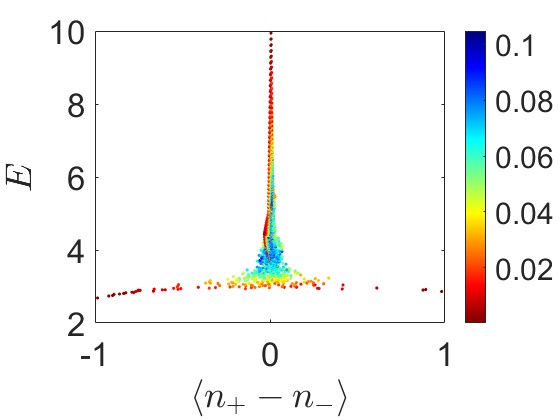}

\includegraphics[width=4cm]{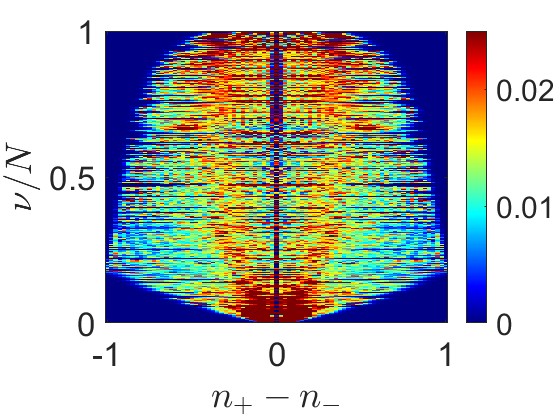}
\includegraphics[width=4cm]{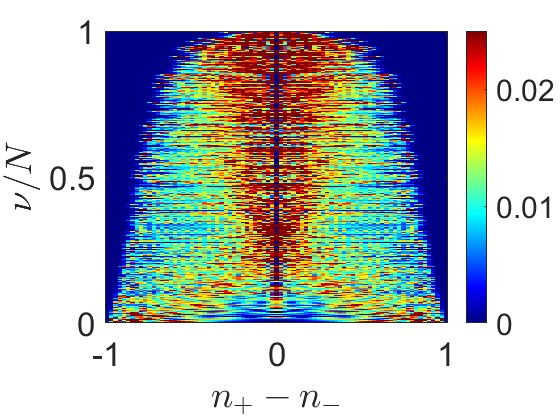}

\caption{
{\bf The spectrum of a rotating trimer.}
The upper left and right panels display the spectra for ${\Phi=0+\delta\Phi}$ and for ${\Phi=3\pi+\delta\Phi}$. The horizontal axis is the calculated expectation value ${\braket{n_{+}{-}n_{-}}}$ for each eigenstate. Due to ${\delta\Phi=0.1\pi}$ the quasi-degenerate ground-states exhibit symmetry breaking. The interaction is ${u=20}$, as in the corresponding panel of \Fig{trimerSpect}.
The lower panels use ``image plot" for the representation of the same spectra. To each eigenstate corresponds a row (rather than a point) that shows the probability distribution of ${(n_{+}{-}n_{-})}$. Here $\delta\Phi=0$, because symmetry breaking is not required for demonstrating the bistability. 
%
}
\label{trimerRotS}  
%
\ \\
%

\includegraphics[width=5cm]{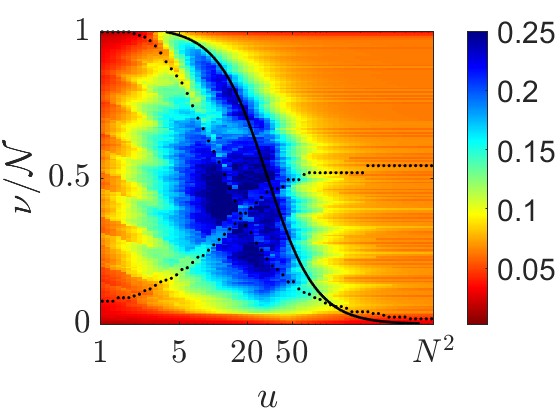}

\caption{
{\bf Phase diagram for a rotating trimer} 
The same as \Fig{trimerTomog}c, but for a rotating trimer with ${\Phi=3\pi}$. Namely, the color indicates $\overline{\mathcal{M}}$. The dotted lines are the borders $E_{SB}$ (ascending) and $E_{SF}$ (descending). The solid line is $u_s(E)$.   
%
}
\label{trimerRotD}  
\end{figure}

\section{Tomography of rotating trimer}
\label{aH}

For a rotating trimer with ${\Phi \sim 3\pi}$ there is a pair of states that form a quasi-degenerate ground-state, one features condensation at the $k=2\pi/3$ orbital, and the other features condensation at the $k=-2\pi/3$ orbital. Formally one of them is the actual ground-state, while the other is a metastable state. A representative spectrum is displayed in \Fig{trimerRotS}.  
The horizontal axis is the average population imbalance ${(n_{+}{-}n_{-})}$ of the $k=\pm 2\pi/3$ orbitals. For sake of comparison we also display the spectrum for ${\Phi \sim 0}$ where the ground date is condensation at the ${k=0}$ orbital. 
In the lower panels of \Fig{trimerRotS} we display ``image plot" for the representation of the same spectra. The symmetry breaking is clearly reflected in the probability distribution of the ${\Phi = 3\pi}$ eigen-functions.  

In \Fig{trimerRotD} we display the ``standard" $(u,E)$ phase-diagram of the rotating trimer. The lines that we indicate in this diagram are $E_{SB}$ and $E_{SF}$, where 
\beq \nonumber
E_{min} &=& \frac{1}{6}N^2U - \frac{1}{2} NJ, 
\ \ \ \ \ [\text{degenerated}] 
\nonumber \\
E_{SF} &=& \frac{1}{6}N^2U + NJ 
\nonumber \\
E_{SB} &=& \frac{1}{4}N^2U - \frac{1}{2} NJ 
\eeq 
The $E_{min}$ and $E_{SF}$ expressions are swapped versions of \Eq{eBorders}, due to having ${\Phi = 3\pi}$ instead of ${\Phi = 0}$. Note that $V_{-}(\bm{n})$ has two degenerated minima. The additional threshold $E_{SB}$ is the potential barrier between the two quasi-degenerate minima.  Its determination is analogous to $E_{ST}$, but now the focus is on the {\em lower} potential surface. For $E<E_{SB}$ the accessible $\bm{n}$-space is fragmented into two disconnected branches, hence symmetry breaking is observed.

\section{Outstanding MI states}
\label{aI}

Superpositions of Fock-states that have the same interaction energy may lead to non-zero $n_0$ indicating ODLRO if they  involve configurations that differ by one-particle transition, and outstanding value of $\beta$ if they involve  configurations that differ by at most two particle transitions. Here we provides details of the calculation, focusing on the first excitation band of trimer.        

The generic value for $n_0$ is $\bar{n}=N/3$. This is what we get for any definite Fock occupation state of the trimer. But if we have a superposition $\Psi$ os Fock states we have to calculate the following enhancement factor
\beq
\alpha_1 \ = \ \frac{\braket{\hat{n}_0}}{\braket{\hat{n}_0}_g} \ = \ 
\frac{\BraKet{\Psi}{\sum_{i,j}a_i^{\dag}a_j}{\Psi}}{\BraKet{\Psi}{\sum_j a_j^{\dag}a_j}{\Psi}} 
\eeq
where the subscript ``g" stands for ``generic".
In the generic case this factor equals unity because the transitions $a_i^{\dag}a_j$ (with ${i \ne j}$) give zero contribution. Let us consider an example where this is not the case. The first band of excitations is spanned by 6 basis states, namely, 
\beq \label{eFockS}
\ket{1a} &=& \ket{\bar{n}{-}1,\,\bar{n}{+}1,\,\bar{n}} \\ 
\ket{1b} &=& \ket{\bar{n}{-}1,\,\bar{n},\,\bar{n}{+}1} \\ 
\ket{2a} &=& \ket{\bar{n},\,\bar{n}{-}1,\,\bar{n}{+}1} \\ 
\ket{2b} &=& \ket{\bar{n}{+}1,\,\bar{n}{-}1,\,\bar{n}} \\ 
\ket{3a} &=& \ket{\bar{n}{+}1,\,\bar{n},\,\bar{n}{-}1} \\ 
\ket{3b} &=& \ket{\bar{n},\,\bar{n}{+}1,\,\bar{n}{-}1} 
\eeq
With one-particle-transition we can connect any of this states to two other states in the same set. The lowest excitation is the superposition 
\beq \label{ePsiH}
\ket{\Psi} \ \ = \ \ \frac{1}{\sqrt{6}} \sum_{x,s} \ket{x,s} 
\eeq
It is clear that for this superposition, due to the 2 extra transitions per basis state, the enhancement factor is $\alpha_1=5/3$, hence ${n_0=(5/9)N}$.

We now discuss the enhancement factor for the second moment calculation. Based on \Eq{enocc} we get:
\beq
\alpha_2=\frac{ \braket{(\hat{n}_0-\bar{n})^2}}{\braket{(\hat{n}_0-\bar{n})^2}_g} 
= 
\frac
{\BraKet{\Psi}{\sum' (a_{i'}^{\dag}a_{j'}) (a_j^{\dag}a_i) }{\Psi}} 
{\BraKet{\Psi}{\sum' (a_i^{\dag}a_j) (a_j^{\dag}a_i) }{\Psi}} 
\ \ \ 
\eeq
where the prime indicates summation over all the transitions with ${i \ne j}$ and ${i' \ne j'}$.  For the $\Psi$ of \Eq{ePsiH} we have in the denominator the 6 generic contributing terms, while in the numerator one realize that there are 15 contributing terms. Accordingly the enhancement factor is ${\alpha_2=15/6=5/2}$. 

The present example is somewhat complicated because both the first and the second moments of $\hat{n}_0$ are non-generic. Accordingly, the enhancement factor of the variance is    
\beq
\alpha_0 = \frac{\text{Var}(n_0)}{\text{Var}(n_0)_g}  
= \alpha_2 - (\alpha_1{-}1)^2 
\frac{\bar{n}^2}{\braket{(\hat{n}_0-\bar{n})^2}_g}   
\eeq
Based on \Eq{eSigG} we have 
$\braket{(\hat{n}_0-\bar{n})^2}_g=(2/27)N^2$, 
and accordingly $\alpha_0=11/6$.

The extra polarization also affect $\sigma_{\perp}$. Namely, the sum rule \Eq{eSRt} implies that $\sigma_{\perp}$ is suppressed by a factor ${\alpha_{\perp}=8/9}$. Accordingly the enhancement factor of $\beta$ is $\alpha_0/\alpha_{\perp}=33/16$. Thus we get the outstanding value ${\beta=11/8}$. 

The upper excitation band that is spanned by ${\ket{N{-}1,\, 1, \, 0} }$ and its permutations can be treated in a similar way. The calculation is somewhat simpler because the extra-polarization $\braket{(\hat{n}_0-\bar{n})}$ is negligible. For nearby non-generic bands, it is in fact strictly zero. Therefor only $\alpha_2$ is significant for the non-generic $\beta$ calculation.  Note however that in the calculation of this enhancement factor, the different transitions have different weights because the occupation is not uniform.

\clearpage

\clearpage
\end{document}